\documentclass[fleqn,usenatbib]{mnras}

\usepackage[T1]{fontenc}
\usepackage{ae,aecompl}
\usepackage{mathtools}

\usepackage{graphicx}	
\usepackage{amsmath}	
\usepackage{amssymb}	
\usepackage{siunitx}
\usepackage{comment}
\usepackage{mwe}
\usepackage{subfig}
\usepackage{dblfloatfix}
\usepackage[percent]{overpic}
\usepackage[usenames,dvipsnames]{xcolor}

\newcommand{\ds}{$\, {\rm deg}^2$}
\newcommand{\angstrom}{\text{\normalfont\AA}}

\usepackage{newtxtext,newtxmath}

\title[The rest-frame UV LF at $3<z<5$]{The total rest-frame UV luminosity function from $3 < z < 5$: A simultaneous study of AGN and galaxies from $-28<M_{\rm UV}<-16$}

\author[N. J. Adams et al.]{
N. J. Adams$^{1,2}$\thanks{E-mail: nathan.adams@manchester.ac.uk},
R. A. A. Bowler$^{1,2}$,
M. J. Jarvis$^{2,3}$,
R. G. Varadaraj$^{2}$,
B. H\"au\ss ler$^{4}$,
\\
$^{1}$Jodrell Bank Centre for Astrophysics, University of Manchester, Oxford Road, Manchester, UK\\
$^{2}$Sub-department of Astrophysics, University of Oxford, Denys Wilkinson Building, Keble Road, Oxford OX1 2DL, UK\\
$^{4}$Department of Physics, University of the Western Cape, Bellville 7535, South Africa\\
$^{5}$European Southern Observatory, Alonso de Cordova 3107, Vitacura, Santiago, Chile\\}

\pubyear{2022}

\begin{document}
\label{firstpage}
\pagerange{\pageref{firstpage}--\pageref{lastpage}}
\maketitle

\begin{abstract}
We present measurements of the rest-frame ultraviolet luminosity function at redshifts $z=3$, $z=4$ and $z=5$, using 96894, 38655 and 7571 sources respectively to map the transition between AGN and galaxy-dominated ultraviolet emission shortly after the epoch of reionization. Sources are selected using a comprehensive photometric redshift approach, using $10$\ds\, of deep extragalactic legacy fields covered by both HSC and VISTA. The use of template fitting spanning a wavelength range of $0.3\text{--}2.4\mu$m achieves  $80\text{--}90$ per cent completeness, much higher than classical colour-colour cut methodology. The measured LF encompasses $-26<M_{\rm UV}<-19.25$. This is further extended to $-28.5<M_{\rm UV}<-16$ using complementary results from other studies, allowing for the simultaneous fitting of the combined AGN and galaxy LF. We find that there are fewer UV luminous galaxies ($M_{\rm UV}<-22$) at $z\sim3$ than $z\sim4$, indicative of an onset of widespread quenching alongside dust obscuration, and that the evolution of the AGN LF is very rapid, with their number density rising by around 2 orders of magnitude from $3<z<6$.  It remains difficult to determine if a double power law (DPL) functional form is preferred over the Schechter function to describe the galaxy UV LF. Estimating the Hydrogen ionizing photon budget from our UV LFs, we find that AGN can contribute to, but cannot solely maintain, the reionization of the Universe at $z=3-5$. However, the rapidly evolving AGN LF strongly disfavours a significant contribution within the EoR.

\end{abstract}

\begin{keywords}
galaxies: evolution -- galaxies: formation -- galaxies: high-redshift
\end{keywords}



\section{Introduction}

Over the past 30 years, survey programmes exploiting various combinations of area and depth have allowed for the rest-frame ultraviolet (UV, 1500\angstrom) luminosity function (LF) of galaxies and active galactic nuclei (AGN) to be measured across a wide range of luminosities and cosmic time. However, at high redshifts ($z>3$) there are many outstanding questions on the shape and evolution of the UV LF. The luminosity regime $-24<M_{\rm UV}<-22.5$ is especially challenging, as this is where the galaxy and AGN comoving space densities become comparable \citep{Ono2017,Stevans2018,Adams2020,Bowler2021,Harikane2021}. Here, the space density is low enough that \emph{Hubble} Space Telescope (HST) programmes do not provide enough cosmic volume to produce statistically significant samples of objects \citep[e.g.][]{Oesch2010,McLure2013,Finkelstein2015,Bouwens2015,Parsa2016,Ishigaki2018,Bouwens2021}. Meanwhile, ground-based observations cover larger survey volumes, but lower resolution makes morphological differentiation between galaxies and AGN \citep[e.g.][]{Masters2012,Akiyama2018,Matsuoka2018c,Niida2020} more challenging. In parallel, this redshift/luminosity regime is also often lacking in spectroscopic completeness over the survey volumes required to characterise the objects as galaxy or AGN \citep[e.g.][]{Ikeda2012,McGreer2013,McGreer2018,Ono2017}. Ultimately, this makes it difficult to accurately separate the two populations.

These issues of selection and completeness result in contrasting conclusions regarding the faint end of the AGN luminosity function. For example, derived faint-end slopes range from $-2.1<\alpha_{AGN}<-1.3$ at $z\sim4$ \citep{Akiyama2018,Stevans2018,Adams2020}. In addition, there is an ongoing debate over whether the bright end of the galaxy population is better described with a double power law (DPL) or Schechter function \citep{Schechter1976} beyond some characteristic luminosity \citep{Bowler2015,Ono2017,Stevans2018,Adams2020,Bowler2020,Harikane2021} and to what degree lensing effects influence bright-end measurements \citep[e.g.][]{Mason2015,Ono2017,Bowler2020}. The primary difference between the DPL and Schechter functions is the gradient of the slope brightwards of the characteristic luminosity ($L^*$) or `knee', leading to differing measurements for the comoving space density of the most luminous systems. An excess in the number density of ultra-luminous galaxies $M_{\rm UV}<-23$ has implications for the efficiency of quenching mechanisms at high redshifts, as well as the impact of dust attenuation \citep[e.g.][]{Bouwens2008,Somerville2012,Cai2014,Ono2017}. However, constraining the numbers of these galaxies with observations proves difficult, as the steepness of the bright-end slope is such that galaxies are quickly outnumbered by AGN-dominated sources at $z\leq6$. A route to solving these problems is to simultaneously fit both the AGN and galaxy UV LF \citep[e.g.][]{Stevans2018,Adams2020,Harikane2021}. This ensures that all sources are accounted for, and that AGN that have significant contributions from their host galaxy are not discounted through various selection criteria, allowing us to consider all sources of UV emission \citep{Bowler2021}. 

Constraining the comoving space density of highly luminous galaxies and UV-faint AGN are key to understanding several other issues in extragalactic astronomy at high redshifts. The first of these is the contribution of AGN towards the budget of ionizing photons in the latter stages of reionization, where there has been significant debate regarding whether AGN are able to sustain the reionization process \citep[e.g.][]{Madau2015,Yoshiura2017,Bosch2018,Hassan2018,Parsa2018,Dayal2020}. Recent measurements have shown that there is a sudden rise in AGN number density post-reionization $3<z<6$ \citep{McGreer2013,Jiang2016,McGreer2018,Kulkarni2019,Kim2020,Niida2020,Zhang2021}, which could have implications for the true numbers of AGN present in the Era of Reionization itself ($z>6$). In parallel, there is a large uncertainty in the growth rates of black holes and AGN activity timescales in the early Universe, leading to multiple theoretical formation mechanisms of the first supermassive black holes \citep[e.g.][]{Banados2018,Wang2019,Yang2020,Wang2021}. Finally, the advent of both deep and wide datasets from the likes of the Subaru and VISTA telescopes has generated debate regarding growth rates of the most luminous and massive galaxies during early times and the impact of dust on their ultraviolet emission \citep[e.g.][]{Ouchi2009,Bowler2017,stefanon2019,Bowler2020,Forrest2020,Neeleman2020}.

Many recent measurements of the UV LF probing the redshift range of $3<z<5$ have made use of colour-colour cuts in order to select galaxy samples \citep[e.g.][]{vdb2010,Bouwens2015,Ono2017,Harikane2021}. While colour-colour cuts can be effective in selecting high-redshift samples using the Lyman break at $\lambda_{\rm rest} < 1216$\AA\, \citep{Guhathakurta1990,Steidel1992,Steidel1996}, there are significant compromises that must be made regarding completeness and contamination rates. This is because regions of colour-colour space are shared with lower redshift galaxies and Milky Way brown dwarf stars \citep{Stanway2008,Bowler2014,Wilkins2014}. A balance is thus required to obtain high sample completeness while minimising contamination. Studies have shown that with careful selection, sample completeness with colour-colour cuts can average 60-70 per cent and peak up to 90 per cent \citep[e.g.][]{Ouchi2004,vdb2010,Ono2017,Harikane2021,Harikane2022}. Extensive near-infrared observations that complement optical observations provide the opportunity to improve upon selection criteria by leveraging the full SED of each source rather than focusing on a select few optical colours. Studies utilising SED modelling and photometric redshifts have shown that completeness can be as high as $\sim80\text{--}90$ per cent in these redshift ranges, while minimising the potential for contamination \citep[e.g.][]{Mclure2009,Bowler2015,Stevans2018,Adams2020}.

In this study, we measure the UV LF consistently across three redshift bins: $2.75<z<3.5$, $3.5<z<4.5$ and $4.5<z<5.2$ using $\sim10$\ds\, of sky containing deep photometry in both the optical and near-infrared. This is performed with the aim of constraining the comoving space density of UV-faint AGN and UV-luminous galaxies, in order to better understand the rise in AGN after reionization and determine if the bright end of the galaxy population can be better described with a Schechter function or DPL functional form. We conduct this measurement with the use of deep optical and near-infrared photometry available from HyperSuprimeCam and the VISTA telescope in the legacy science fields of COSMOS, XMM-LSS and Extended Chandra Deep Field South (E-CDFS). The surveys covering these fields provide an ideal depth-area combination, enabling rare sources as bright as $M_{\rm UV}=-26$ to be detected while being simultaneously deep enough to obtain complete samples of sources as faint as $M_{\rm UV}\sim-20.5$ at $z\sim5$. We combine these measurements with results of studies that use other datasets, whose depth-area combinations are more optimised for exploring brighter and fainter luminosities, to expand the luminosity range to $-28.5<M_{\rm UV}<-16$ and enable for the simultaneous modelling of the entire observable luminosity function at these redshifts.

In Section~\ref{sec:data}, we describe the optical and NIR photometric data used in this study and how this data is used to estimate photometric redshifts for sample selection. In Section~\ref{sec:method} we assess the completeness of the sample, and then measure and fit the UV LF. In Section~\ref{sec:LFs}, we detail the results of our fitting procedure and comment on the performance of the different models. In Section~\ref{sec:discussion}, we expand discussion to contextualise our results with the evolution of galaxies and AGN. We present our conclusions in Section~\ref{sec:conclusions}.
Throughout this work, we assume a standard cosmology with $H_0=70$\,km\,s$^{-1}$\,Mpc$^{-1}$, $\Omega_{\rm M}=0.3$ and $\Omega_{\Lambda} = 0.7$ to allow for ease of comparison with other LF studies. All magnitudes listed follow the AB magnitude system \citep{Oke1974,Oke1983}.

\section{Data and Sample Selection}\label{sec:data}

\subsection{Photometry}

\begin{figure}
\centering
\includegraphics[width=1.05\columnwidth]{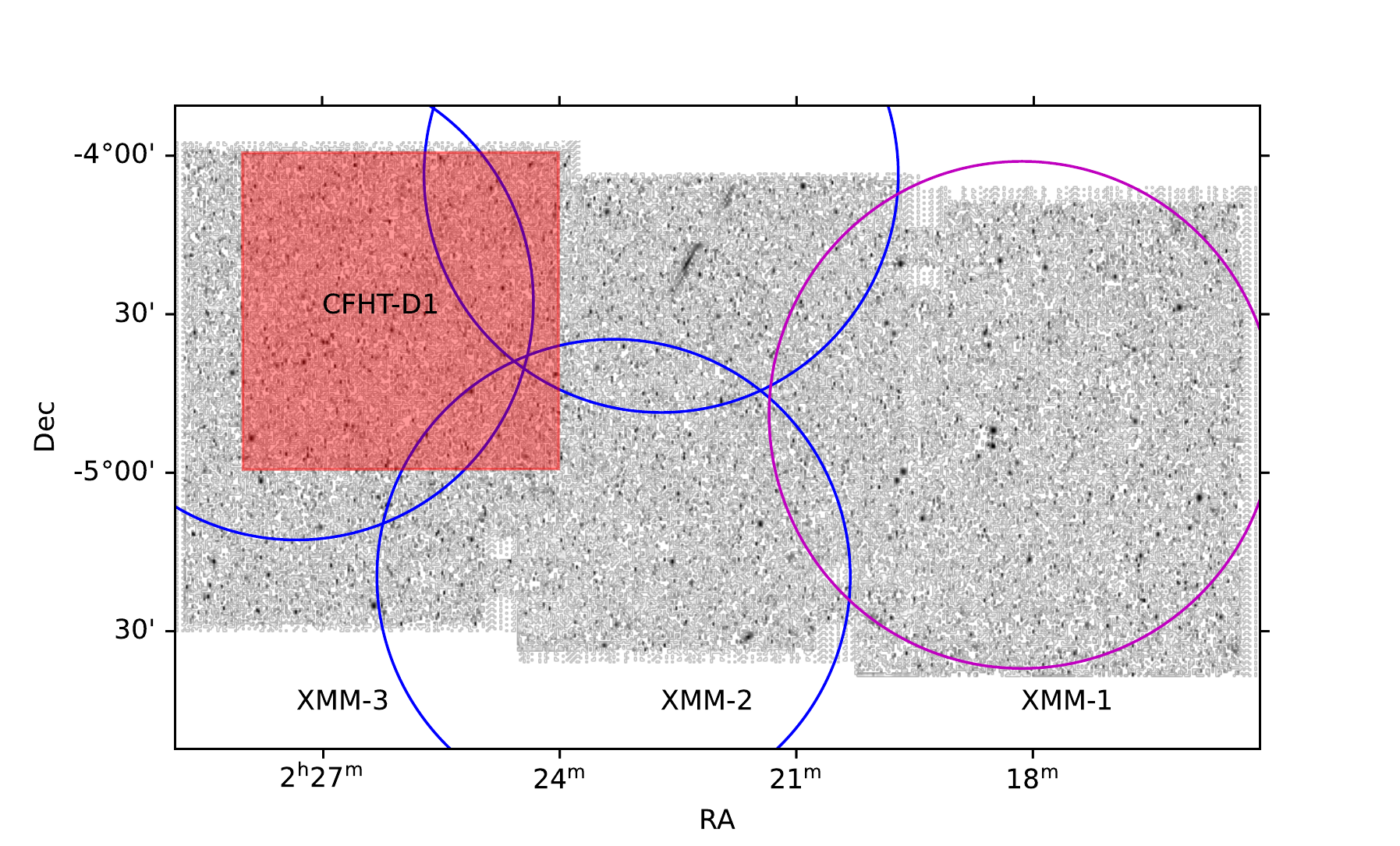}\caption{The footprint coverage of the various surveys in the XMM-LSS field. The background greyscale images are from the VISTA/VIDEO $K_s$-band observations. The VIDEO data products are provided as three primary tiles, which are labelled XMM-1, XMM-2 and XMM-3. The red shaded region indicates the location of the 1\ds\, CFHT-D1 sub-region. The large circles indicate the pointings from HSC DR2, with the blue rings indicating the shallower observations defining our XMM-DEEP sub-region and the magenta circle indicating the deeper HSC observations that define our XMM-UDEEP sub-region.}
\label{fig:XMMField}
\end{figure}

\begin{figure}
\hspace{-0.7cm}
\includegraphics[width=1.1\columnwidth]{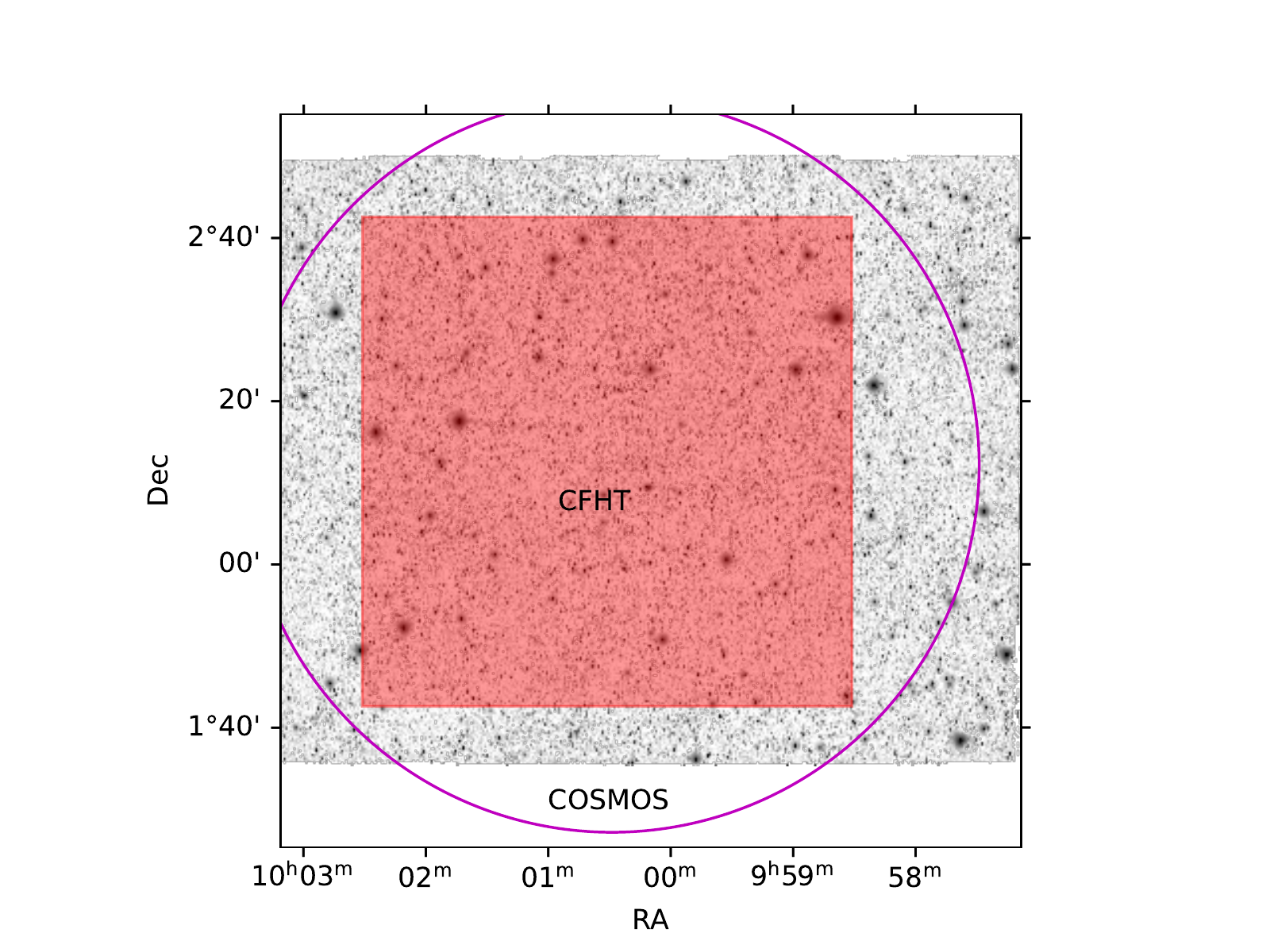}\caption{The footprint coverage of the various surveys in the COSMOS field. The background greyscale images are from the UltraVISTA $K_s$-band observations. The red shaded region indicates the location of the 1\ds\, region containing CFHT coverage. The large circle indicates the deep pointing from HSC DR2. The overlap between HSC and UltraVISTA is used to define the COSMOS sub-region used in this study.}
\label{fig:COSField}
\end{figure}

\begin{figure}
\hspace{-0.6cm}
\includegraphics[width=1.2\columnwidth, height=0.95\columnwidth]{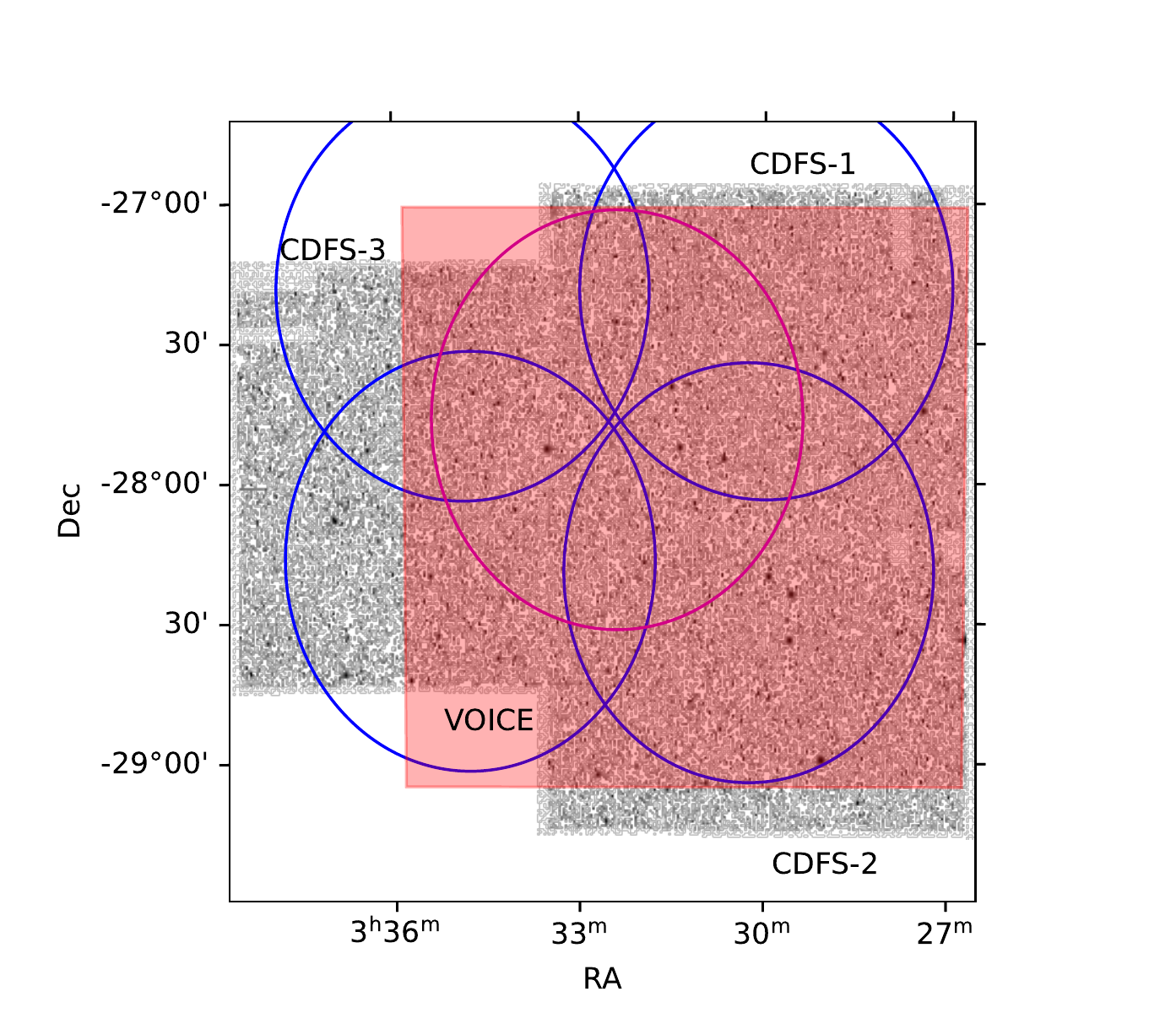}\caption{The footprint coverage of the various surveys in the E-CDFS field. The background greyscale images are from the VISTA/VIDEO $K_s$-band observations. The VIDEO data products are provided as three primary tiles, which are labelled CDFS-1, CDFS-2 and CDFS-3. The red shaded region indicates the location of the VST/VOICE optical coverage. The large circles indicate the pointings from HSC \citep{Ni2019}. In this field, the central magenta pointing is the only pointing from HSC containing $r$-band data.}
\label{fig:CDFSField}
\end{figure}

In this study, we use a multiwavelength dataset spanning up to 14 photometric bands covering $0.3\text{--}2.4\mu$m across three extragalactic fields (XMM-LSS, COSMOS, E-CDFS). Compared to the work in \citet{Adams2020}, the new addition of E-CDFS increases the total area used by $\approx 60$ per cent. Measurements in the photometric bands are derived from the Canada-France-Hawaii-Telescope Legacy Survey \citep[CFHTLS;][]{Cuillandre2012}, VST Optical Imaging of the CDFS and ELAIS-S1 Fields \citep[VOICE;][]{Vaccari2016} and the HyperSuprimeCam Strategic Survey Programme \citep[HSC DR2;][]{Aihara2014,Aihara2017,Aihara2019,Ni2019} in the optical regime. Near-infrared data is provided by the final data release of the VISTA Deep Extragalactic Observations (VIDEO) survey \citep{Jarvis2013} for XMM-LSS and CDFS, while UltraVISTA DR4 \citep{McCracken2012} provides the near-infrared coverage in COSMOS. A breakdown of the available filters and the depths within each subfield is provided in Table~\ref{tab:FiveSig}. Additionally, a visual representation of the coverage provided by each survey in each field is provided in Figures \ref{fig:XMMField}, \ref{fig:COSField} and \ref{fig:CDFSField} for XMM-LSS, COSMOS and E-CDFS respectively.

With the optical data utilised in this study, the three primary extragalactic fields can be broken down into a total of 5 sub-regions containing approximately uniform depth in the imaging. We designate these regions COSMOS (COS), XMM-Deep (XMMD), XMM-UltraDeep (XMMU), CFHT-D1 (D1) and Chandra Deep Field South (E-CDFS). The XMMU sub-region is centred on the Ultra Deep Survey from the UK Infrared Telescope \citep[UKIDSS;][]{Lawrence2007}. Coverage in the $u^*$-band in XMM-LSS is obtained from the CFHTLS Deep Field programme where available, and from the Wide Field programme otherwise. We also include the other optical bands from CFHT within the CFHT D1 and D2 fields (See Figures \ref{fig:XMMField} and \ref{fig:COSField}), each covering 1\ds\, within XMM-Deep and COSMOS. Photometry was carried out using SExtractor \citep{Bertin1996} in dual-image mode, using the deepest $r$, $i$, and $z$ photometric bands in each sub-field to capture the rest-frame ultraviolet emission for redshifts 3, 4 and 5 respectively. Flux is measured in 2\,\arcsec \, diameter circular apertures in all bands. These fluxes are then corrected with a point-source aperture correction, where we model the point spread function (PSF) with {\sc PSFEx} \citep{Bertin2011}; see \citet{Bowler2020} for further details on catalogue generation. The images used are seeing limited and have a full width at half maximum of around 0.8-0.9 arcseconds. The sizes of the galaxies we study in this work are sufficiently small ($\sim 1-2kpc$; \citet{Huang2013,Allen2017,Bowler2017}) to be dominated by the PSF size, and hence we are able to measure accurate colours without the need for PSF matched photometry.

The sky area used for each subfield, and subsequently the cosmic volumes probed, are calculated based on the areas in which the optical and near-infrared observations overlap and large artefacts (e.g. stellar ghosting) have been masked out. We present a visual diagram of the layout of the data for each of the three primary fields in Fig. \ref{fig:XMMField} for the XMM-LSS field, Fig. \ref{fig:COSField} for the COSMOS field and Fig. \ref{fig:CDFSField} for E-CDFS. Between the different filters used for source selection in our catalogues, there is only one primary difference in the total areas covered. This difference originates in the E-CDFS field, where the $r$-selected catalogue uses the VOICE+VIDEO footprint, while the $i$ \& $z$-selected catalogues are confined to the region where both VOICE and HSC are present. This is because the HSC data is deeper than VOICE in these two selection bands, but the HSC observations lack $r$-band coverage required to confirm the presence of the Lyman break, requiring VOICE also be included. This results in a loss in a small fraction of total area used in the field ($3.89$ \ds\, compared to 4.08 in the r-band). To summarise the area usage in the other fields, the COSMOS region has a survey area of 1.51\ds\ while the XMMD region is 1.56\ds, XMMU is 1.83\ds\, and CFHT-D1 is 0.89\ds. Combined, these fields provide a total area of 9.68\ds\, for the $i$-band and $z$-band selected catalogues, while the total area is 9.87\ds\, for the $r$-band selected catalogue. 

\begin{table*}
    \centering
      \caption{Summary of the $5\sigma$ detection depths within the COSMOS, XMM-LSS and CDFS fields. The first section of rows provide depths calculated in 2\arcsec ~diameter circular apertures, placed on empty regions of the image. The second section provides the derived aperture correction for each region, where the numbers are the fraction of enclosed flux within a 2\arcsec ~diameter aperture. The final row shows the area of each region utilised.
      The XMM-LSS field is split into three regions. XMMU consists of the region with the deeper HSC pointing. Additionally 1\ds\, of COSMOS and the D1 region of XMM-LSS contain coverage from CFHT-griz bands.}
    \label{tab:FiveSig}
    \begin{tabular}{ccccccc}
    \hline
    Filter      & COSMOS  & XMMU    & XMMD & D1           & CDFS     & Origin     \\ \hline
    \textit{Depths} & & & & & & \\
    $u^*$       & $ 27.1$ & $25.7$  & $ 25.7$ & $27.1$ & $ (24.9) $ & CFHT (VST) \\
    $g^*$       & $ 27.3$ & --      & -- & $ 27.4$        & $(25.9)$   & CFHT (VST) \\
    $r^*$       & $ 26.9$ & --      & -- & $ 26.9$        & $(26.0)$   & CFHT (VST) \\
    $i^*$       & $ 26.6$ & --      & -- & $ 26.4$        & $(24.6)$   & CFHT (VST) \\
    $z^*$       & $ 25.5$ & --      & -- & $ 25.4$        & --       & CFHT \\
    $g$         & $ 27.4$ & $ 26.9$ & $ 26.7$ & $ 26.7$        & $25.5 $  & HSC        \\ 
    $r$         & $ 27.1$ & $ 26.4$ & $ 25.9$ & $ 25.9$      & $24.8$   & HSC        \\ 
    $i$         & $ 26.9$ & $ 26.3$ & $ 25.6$  & $ 25.6$     & $24.9$   & HSC        \\
    $z$         & $ 26.5$ & $ 25.7$ & $ 25.4$  & $ 25.4$     & $24.2$   & HSC        \\
    $y$         & $ 25.7$ & $ 24.9$ & $ 24.2$  & $ 24.2$     & --       & HSC        \\
    $Y$         & $25.4$  & $ 25.1$ & $ 25.1$  & $ 25.1$     & $25.1$   & VISTA      \\
    $J$         & $25.3$  & $ 24.7$ & $ 24.7$  & $ 24.7$     & $24.6$   & VISTA      \\
    $H$         & $25.1$  & $ 24.1$ & $ 24.2$  & $ 24.2$     & $24.1$   & VISTA      \\
    $K_{\rm s}$ & $25.0$  & $ 23.8$ & $ 23.9$  & $ 23.9$     & $23.8$   & VISTA      \\ \hline
    \textit{PSF} & & & & & & \\
    $u^*$       & $ 0.83$ & $0.81$  & $ 0.81$ & $ 0.81$ & $ (0.66) $ & CFHT (VST) \\
    $g^*$       & $ 0.84$ & --      & $ 0.82$ & $ 0.82$       & $(0.66)$   & CFHT (VST) \\
    $r^*$       & $ 0.85$ & --      & $ 0.83$ &  $ 0.83$     & $(0.69)$   & CFHT (VST) \\
    $i^*$       & $ 0.85$ & --      & $ 0.85$ & $ 0.85$      & $(0.71)$   & CFHT (VST) \\
    $z^*$       & $ 0.84$ & --      & $ 0.83$ & $ 0.83$      & --       & CFHT \\
    $g$         & $ 0.82$ & $ 0.85$ & $ 0.81$ & $ 0.81$      & $0.66 $  & HSC        \\ 
    $r$         & $ 0.83$ & $ 0.81$ & $ 0.86$ & $ 0.86$      & $0.62$   & HSC        \\ 
    $i$         & $ 0.84$ & $ 0.87$ & $ 0.79$ & $ 0.79$      & $0.83$   & HSC        \\
    $z$         & $ 0.86$ & $ 0.82$ & $ 0.80$ & $ 0.80$      & $0.75$   & HSC        \\
    $y$         & $ 0.79$ & $ 0.83$ & $ 0.75$ & $ 0.75$      & --       & HSC        \\
    $Y$         & $0.72$  & $0.72$ & $0.72$ & $0.72$      & $0.72$   & VISTA      \\
    $J$         & $0.76$  & $0.76$ & $0.76$ & $0.76$      & $0.75$   & VISTA      \\
    $H$         & $0.79$  & $0.78$ & $0.79$ & $0.79$      & $0.78$   & VISTA      \\
    $K_{\rm s}$ & $0.81$  & $ 0.79$ & $ 0.80$ & $ 0.80$      & $0.81$   & VISTA      \\ \hline
     \textit{Area}      & 1.51\ds & 1.82\ds  & 1.56\ds & 0.89\ds & 3.89-4.08\ds & -- \\ \hline
    \end{tabular}
\end{table*}

\subsection{Photometric redshifts}\label{sec:photozs}

\begin{figure*}%
    \centering
    \hspace*{-1.5em}
    \subfloat[XMM-LSS]{
        \begin{overpic}[width=1.04\columnwidth]{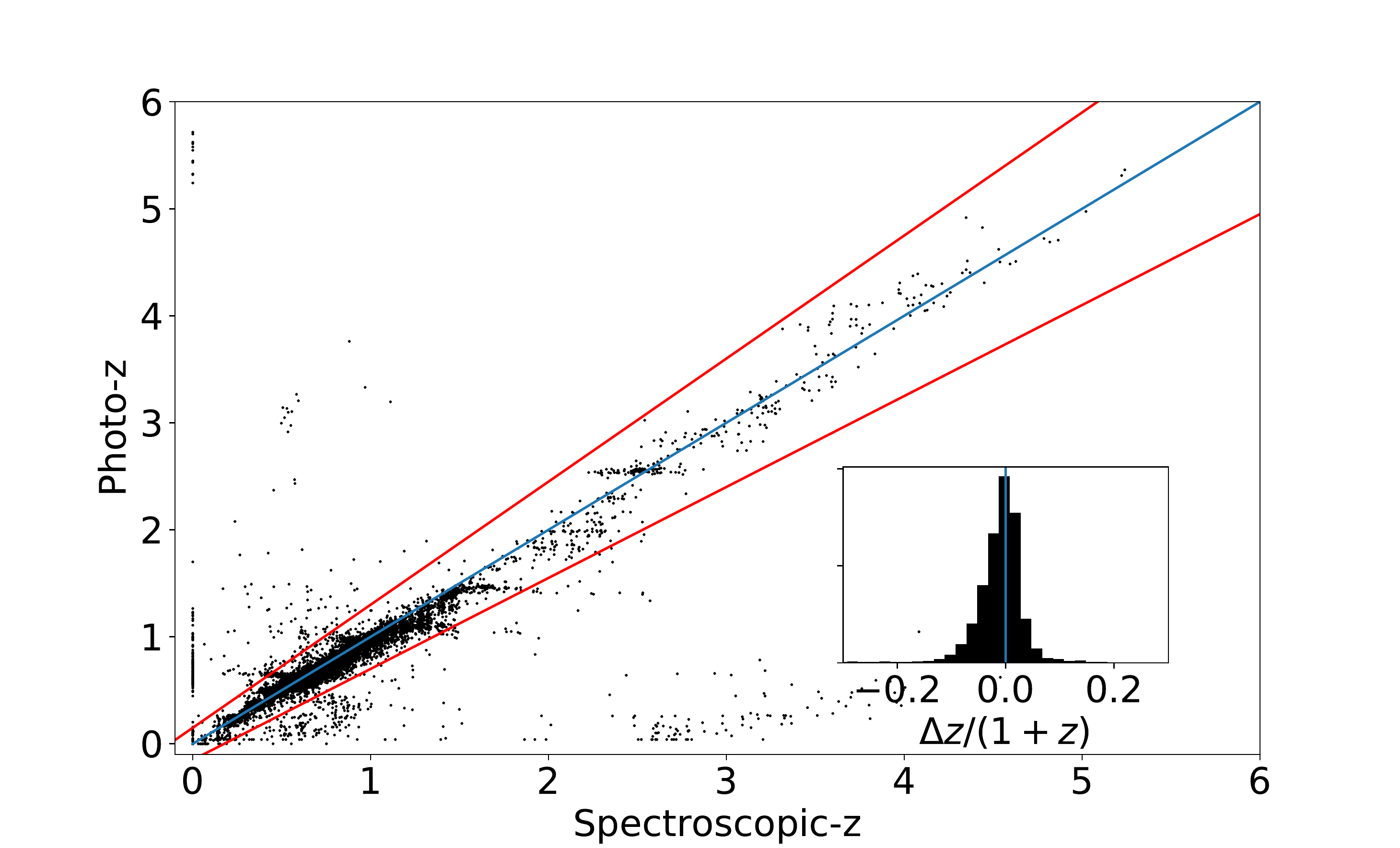}
            \put(15,49){$\eta =4.5 \%$, $\sigma_{\rm NMAD} = 0.031$}
        \end{overpic}
    }%
    \hspace*{-2.9em}%
    \qquad
    \subfloat[COSMOS]{
        \begin{overpic}[width=1.02\columnwidth]{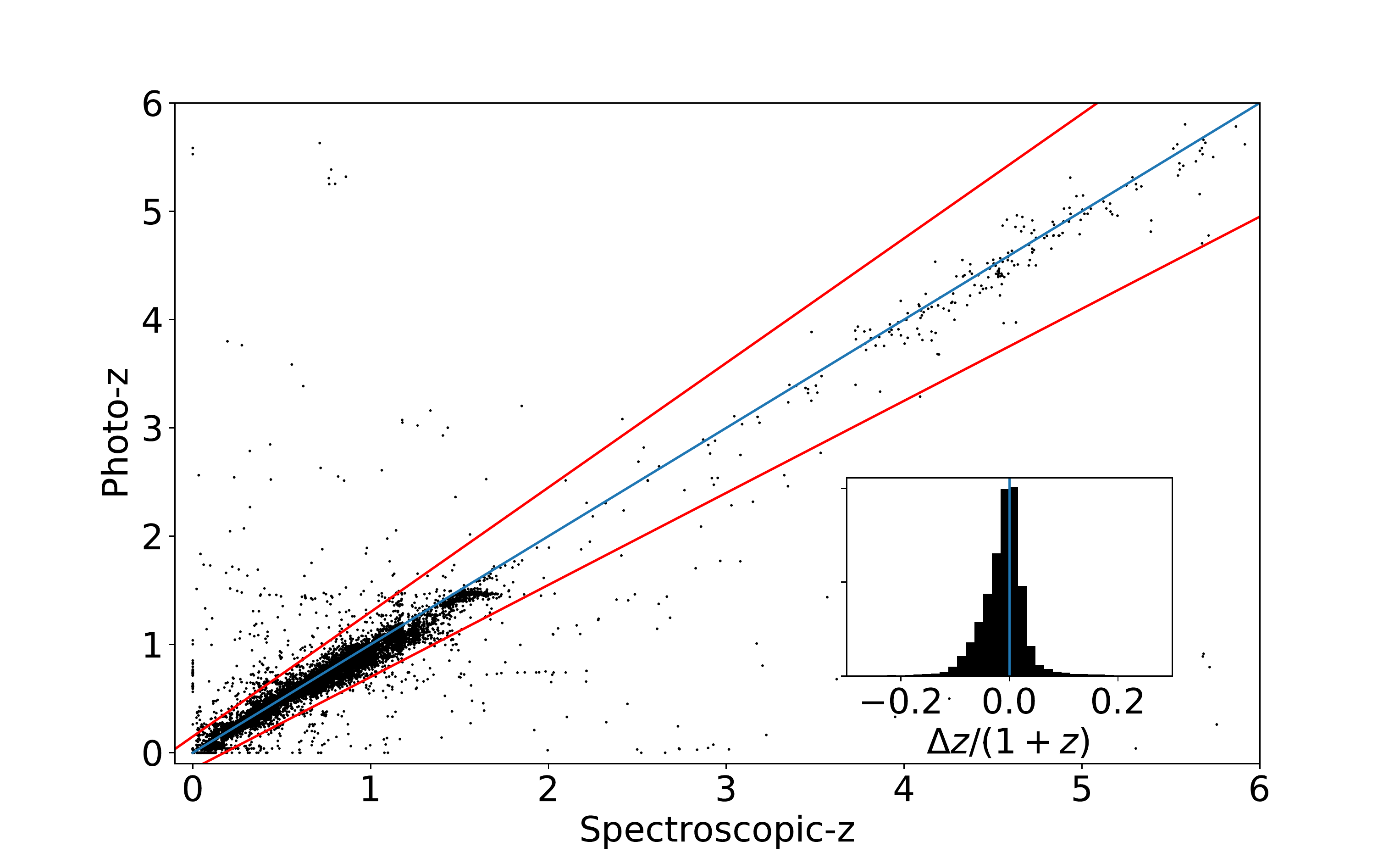}
            \put(15,50){$\eta =3.1 \%$, $\sigma_{\rm NMAD} = 0.029$}
        \end{overpic}
    }%
    \qquad
    \subfloat[E-CDFS]{
        \begin{overpic}[width=1.10\columnwidth]{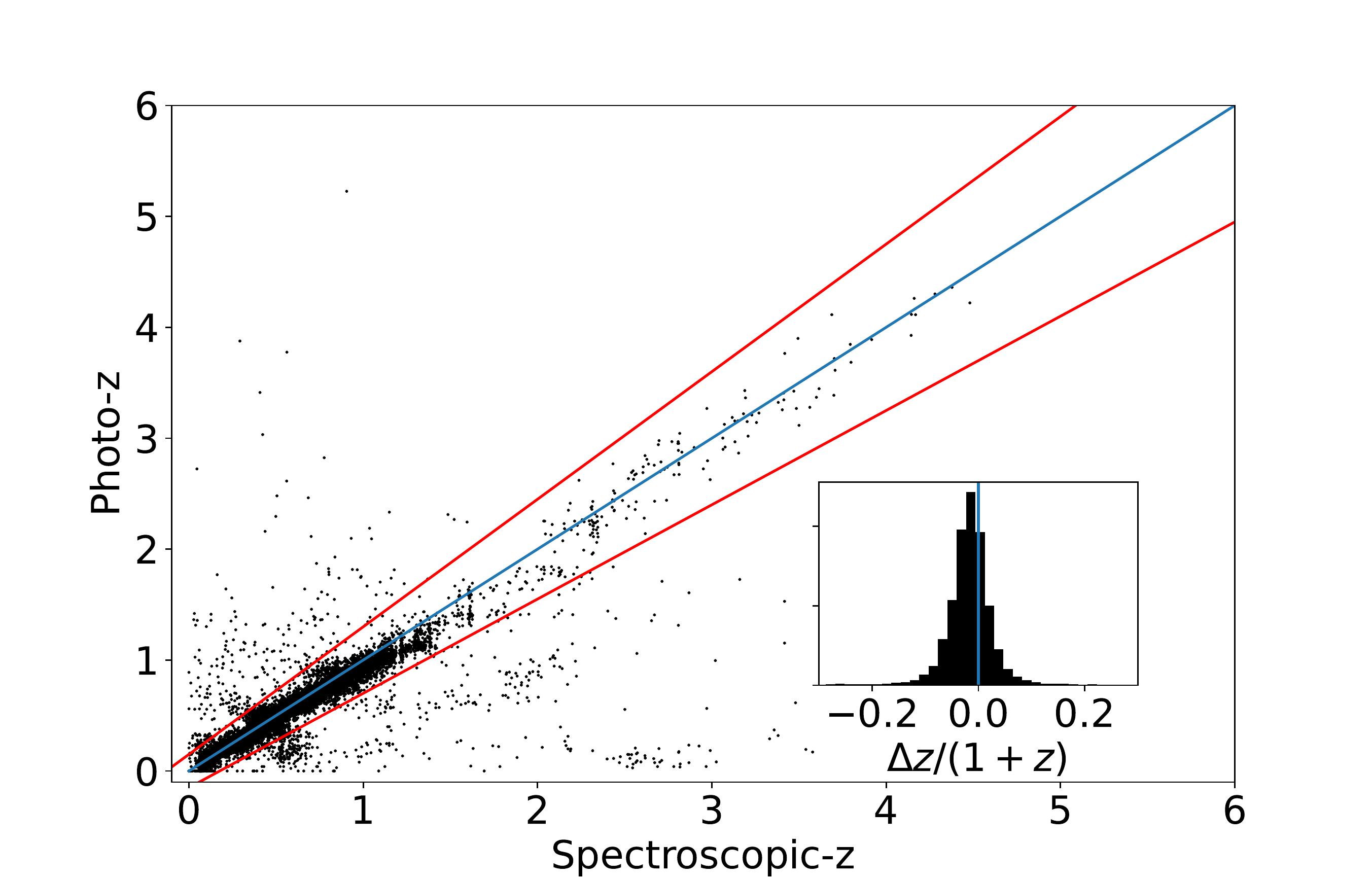}
            \put(15,51){$\eta =4.1 \%$, $\sigma_{\rm NMAD} = 0.039$}
        \end{overpic}
    }%
    \caption{Comparisons between the photometric redshifts derived in this study and a large compilation of spectroscopic redshifts. In panel a) we show the results for the XMM-LSS field, in b) we show COSMOS and in c) we show E-CDFS. The blue line shows the one-to-one correlation in the ideal case, and the red lines define the 15 per cent margin in $1+z$ that defines a significant outlier. The sub-plot in each figure shows the histogram of the scaled photometric redshift deviations from the spectroscopic values. Each figure displays the outlier rate ($\eta$) and NMAD ($\sigma_{\rm NMAD}$) of each sample.}%
    \label{fig:PhotZ}%
\end{figure*}

With such a wealth of multiwavelength data in these extragalactic fields, we select our sample following a spectral energy distribution (SED) template fitting procedure. Object classification and redshift estimates are made using the template fitting photometric redshift code {\sc LePhare} \citep{Arnouts1999,Ilbert2006}. This code minimises the $\chi^2$ of various spectral energy distribution (SED) templates for galaxies, AGN and Milky Way stars using the multi-band photometry and uncertainties. We set the uncertainties of the photometry to a minimum of 5 per cent during the fitting process. This minimum error accounts for potential imperfections in the template sets and the filter transparency functions. It also allows for improved convergence in the model fitting procedure, since we are using a finite set of templates while real galaxies probe a continuous distribution of colours. The template sets used in this study are the COSMOS SED template set for galaxies \citep{Ilbert2009}, AGN/QSO-like objects from \citet{Salvato2009} and stellar templates from a combination of results from \citet{Hamuy1992,Hamuy1994,Bohlin1995,Pickles1998,Chabrier2000}. Additional stellar templates are also added from the SpeX\footnote{https://cass.ucsd.edu/~ajb/browndwarfs/spexprism/index.html} brown dwarf library. These brown dwarf templates were added to the sample because M and L-class dwarfs can have very similar colours to those of $z\sim5$ galaxies (see Section \ref{sec:Sample} for further discussion on brown dwarfs). The treatment derived in \citet{Madau1995} is used for absorption by the inter-galactic medium (IGM) and the \citet{Calzetti2000} dust law is used with varying strengths of $E(B-V) = 0-1.5$.
 
The SED fitting process was conducted in two stages. In the first stage, we cross-match our catalogues to a spectroscopic sample which compiles results from the VVDS \citep{LeFevre2013}, VANDELS \citep{McLure2018,Pentericci2018,Garilli2021}, Z-COSMOS \citep{Lilly2009}, DEIMOS-10K \citep{Hasinger2018}, VIPERS \citep{Scodeggio2018} and FMOS \citep{Silverman2015} surveys. We only use objects with spectroscopic redshifts that have flags indicating a greater than 95 per cent confidence. In addition, the source must have a $5\sigma$ detection in at least one of the detection bands used. This process provides a total spectroscopic sample of 14811 cross-matched sources in XMM-LSS and 18811 sources in COSMOS. Within CDFS, a spectroscopic catalogue of 23947 cross-matched sources was obtained from the Spitzer Data Fusion database \citep{Vaccari2015}, featuring results from surveys including VVDS \citep{LeFevre2013}, BLAST \citep{Eales2009}, ACES \citep{Cooper2012}, GOODS-CANDELS \citep{Hsu2014}, OzDES \citep{Yuan2015,Childress2017} and VUDS \citep{Tasca2017}. We also include new spectroscopic redshifts obtained from the LADUMA collaboration using the 
Anglo-Australian Telescope (Wu et al. in prep). Cross-matched catalogues are then run through {\sc LePhare} in its {\sc AUTO\_ADAPT} mode, which makes iterative adjustments to the zero-points of the photometric filters in order to optimise the results against the spectroscopic sample. This process is carried out separately for each of the three primary fields, leading to three sets of zero point corrections. A diverse spectroscopic sample is required in order to prevent these zero-point modifications from being biased towards a limited set of galaxy colours. The results of this process provides small offsets (mostly 0--0.07mags with the exception of the $u^*$-band at $\sim0.1-0.15$mags) across all bands. These offsets are then applied to the full photometric catalogue and {\sc LePhare} is run a second time to obtain object classification and redshifts for the full sample.
 
Comparisons to the spectroscopic sample can provide indications of the reliability of the photometric redshift estimations. This can be broken down into two numerical values. First, the outlier rate: the fraction of photometric redshifts which disagree with the spectroscopic redshift by more than 15 per cent in ($1+z$), and Second, the Normalised Median Absolute Deviation \citet[NMAD;][]{hoaglin2000understanding}: a measurement of the spread of the photometric redshifts around the ground truth in a manner that is resistant to the few extreme outliers that are present, this is defined as $1.48 \, \times\,$median[|$\Delta z$|/$(1+z)$] where $\Delta z = z_{phot} - z_{spec}$. Across the COSMOS field we find an outlier rate of 3.1 per cent and a NMAD of 0.029, in XMM-LSS the outlier rate is 4.5 per cent and the NMAD is 0.031, and finally CDFS has an outlier rate of 3.5 per cent and NMAD of 0.037. Fig.~\ref{fig:PhotZ} shows the spectroscopic redshifts against our photometric redshift estimates. We note a small bias in the photometric redshifts between the range of $0.8<z<1.4$, where photometric redshifts are systematically lower than spectroscopic redshifts with a mean $\Delta z = -0.06$, three times larger than the rest of the redshift space. This leads to a slight skew in the $\Delta z$ histograms of Fig.~\ref{fig:PhotZ}. However, such a bias is not present when considering the redshift range of interest in this study.

Focusing on the redshift ranges of interest, we find that we successfully recover (within 15 per cent of $1+z$) 592/911 of the cross-matched spectroscopic objects between $2.5<z<3.5$ (65 per cent), 218/257 between $3.5<z<4.5$ (85 per cent) and 125/142 objects between $4.5<z<5.5$ (88 per cent). The majority of cases of incorrectly identified sources at $z\simeq$ 4 and 5 are a result of blending issues that result in a dilution of the Lyman break and an underestimated redshift ($z<1$). Such objects are accounted for as part of our completeness simulations which are detailed in the LF determination (Section~\ref{sec:completeness}). The performance at $z\simeq3$ is worse than that of the other two bins due to the limited wavelength coverage of the study. At these redshifts, the Lyman break at 1216\AA\, lies redwards of the $u$-band, but the level of attenuation bluewards of the break is not as strong as at higher redshifts between the rest frame 912-1216\AA\, \citep{Madau1995,Inoue2014}, leading to a shallower drop in the $u^*$-flux. In addition, the $u^*$-band is often shallower than the $r$-band used for the selection of targets at $z=3$, and a shallow $u^*$-band Lyman break has degeneracies with the Balmer break at $z<0.4$. These effects makes it difficult to ascertain if a target is truly at $2.5<z<3.5$, especially towards the lower redshift boundary, where the influence of the break is at its weakest in the $u^*$-band. It is therefore not a surprise that we find that the spectroscopic redshift recovery rate is found to be around 60 per cent in the redshift region of $2.5<z_{\text{spec}}<2.75$ and increases to 80 per cent at $2.75<z_{\text{spec}}<3.5$, indicating that the issue does indeed lie with sources with the mildest of Lyman breaks in the $u$-band. We therefore limit our $z\sim3$ UV LF to the redshift range $2.75<z<3.5$.

\subsection{Sample selection}\label{sec:Sample}

\subsubsection{Selecting the initial sample of $2.75<z<4.5$ sources}

We select an initial sample of 218,152 and 53,612 sources in the redshift bins of $2.75<z<3.5$ and $3.5<z<4.5$ based on the following criteria. Firstly, the source has a $\geq5\sigma$ detection in the band containing the rest-frame ultraviolet continuum emission ($r$ for $z\simeq3$ and $i$ for $z\simeq4$). Secondly the source has a best-fitting SED template (minimal $\chi^2$) that is a galaxy or a QSO with a redshift in the range of $2.75<z<3.5$ or $3.5<z<4.5$ for the $r$ and $i$ selected samples. A maximum $\chi^2$ cut of 100 is employed in order to remove sources that are possible artefacts or contaminated sources while retaining sources whose statistics might be limited by the use of a discrete template set.

\subsubsection{Selecting the initial sample of $4.5<z<5.2$ sources}

\begin{figure}
\centering
\includegraphics[width=1.05\columnwidth]{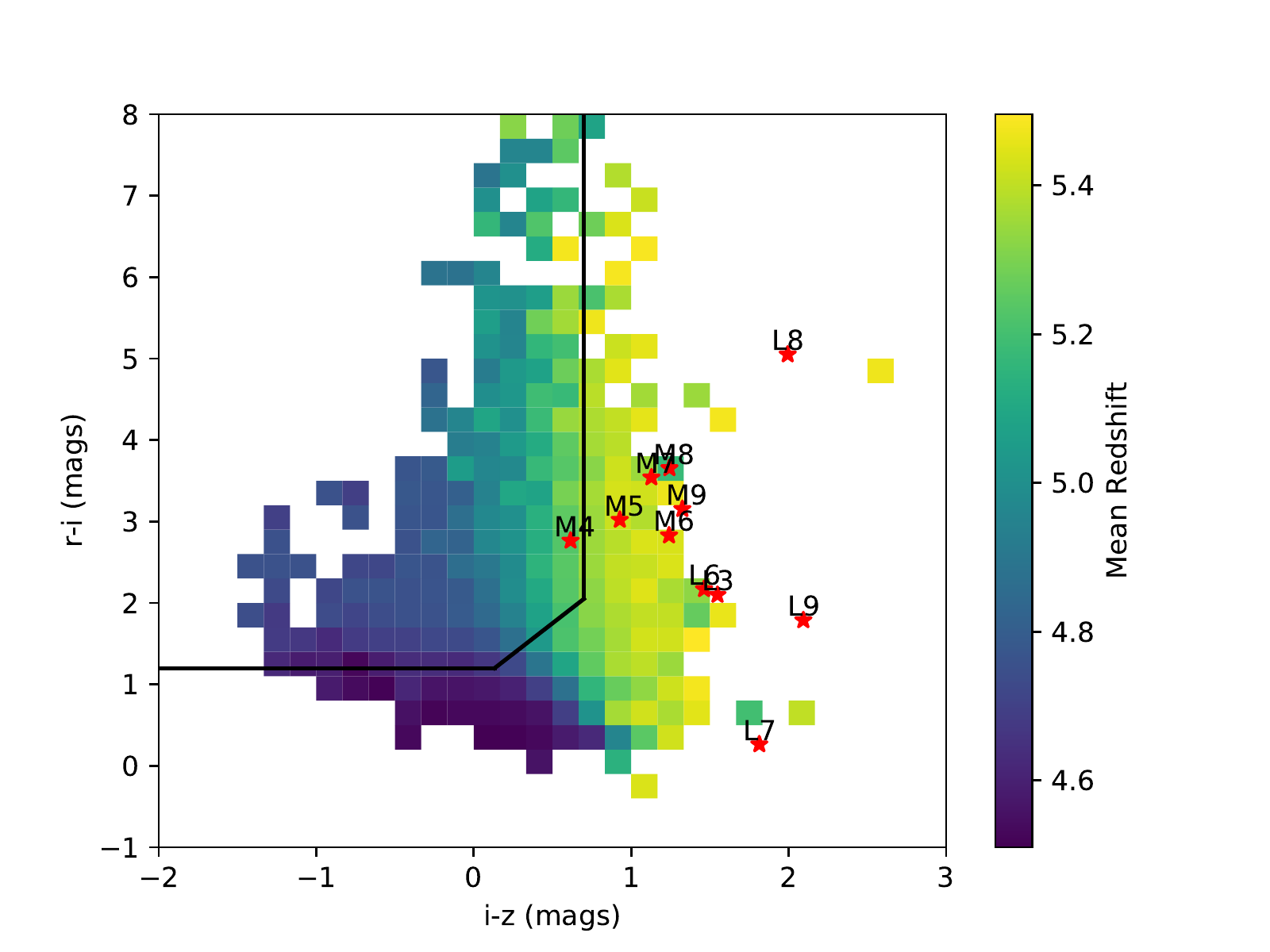}\caption{A colour-colour diagram showing the initial $4.5<z<5.5$ photometric redshift sample around the photometric bands that span the Lyman Break. We group the sample into a 30x30 grid to show how the mean redshift changes across the colour space. Overplotted in solid black lines is the selection criteria employed by \citet{Ono2017} to select galaxies at $z\sim5$ from the region bound by the upper left quadrant. We also show the colours of typical M and L-class brown dwarf stars as the red points and highlight that they overlap with the colour space probed by galaxies with redshifts towards the upper limits ($z>5.2$).}
\label{fig:BD}
\end{figure}

To produce a sample of $z \simeq 5$ galaxies and AGN, we first select 22,489 objects which have a best-fitting SED (minimum $\chi^2$) as being a galaxy or QSO within the redshift range of $4.5 < z < 5.5$ and have a $5\sigma$ detection in the HSC-z band. The strong Lyman break exhibited by galaxies at these redshifts should result in a non-detection in the CFHT-$u^*$ band. As a result, we also implement a requirement for a $<3\sigma$ $u^*$ detection in order to minimise the potential for lower redshift contaminants. However, it was immediately obvious that these cuts alone were insufficient to produce a robust sample of $z\simeq5$ galaxies. Examining the redshift distribution of the sample revealed a large spike in number counts for luminous objects ($M_{\rm UV}<-22$) with $z>5.2$. Such a spike in number counts can be attributed to Milky Way brown dwarfs, particularly M-class dwarfs whose optical/NIR colours become degenerate with high-redshift galaxies at $z>5.2$ (see Fig.~\ref{fig:BD} for an example of the colours of $z\sim5$ galaxies and brown dwarf stars around the redshifted Lyman Break). The inclusion of the brown dwarf templates from the SpeX dataset was found to reduce the number of ultra-luminous sources at $z>5.2$ by greater than a factor of two, but a significant spike in apparently luminous objects remained. 

A possible solution to this problem is the inclusion of \emph{Spitzer} IRAC data, which can break the degeneracy between $z\sim5$ galaxies and M-class brown dwarf stars. This is because the brown dwarfs broadly follow Planck's law and decrease in luminosity towards the mid-infrared, while high-redshift galaxies remain flat \citep[See][ for more details]{Bowler2014}. However, deep \emph{Spitzer} data has a larger point spread function and the subsequent source blending issues can introduce an additional layer of complexity into the sample selection procedure. A second solution would be to introduce an upper limit of $z<5.2$ to the UV LF. Such a cut only causes a small shift in the mean redshift of the UV LF from $\overline{z}=4.9$ to $\overline{z}=4.8$ and greatly minimises the overlap in colour space between galaxies and brown dwarfs. We thus proceed with restricting the sample to $4.5<z<5.2$ and again apply a quality cut of $\chi_{\text{best}}<100$, which removes the worst fit $1$ per cent of objects from the sample. This process provides a total of 15,025 galaxy and AGN candidates between $4.5<z<5.2$ with which we can measure the UV LF.

\begin{figure*}
\centering
\includegraphics[width=1.0\textwidth]{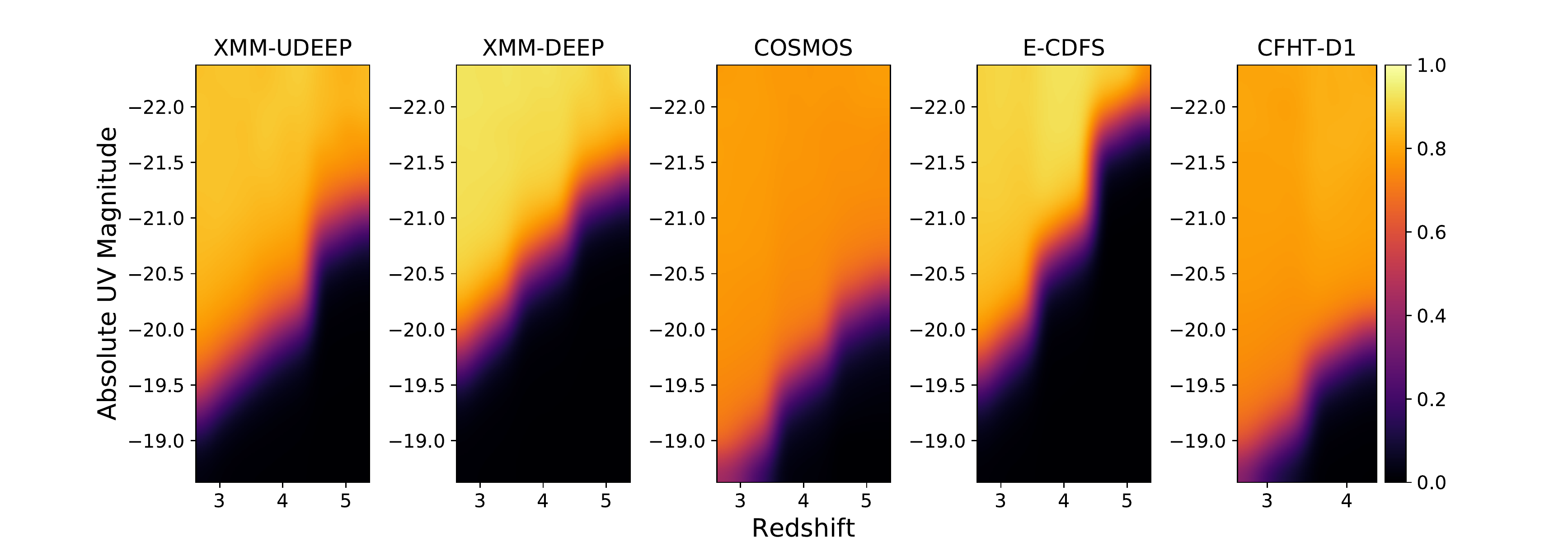}\caption{The completeness distribution as derived from the injection + recovery simulation performed in each of the five sub-fields. The Y-axis displays the rest-frame ultraviolet magnitude of the simulated source while the X-axis displays the input redshift. The lighter (upper, left) regions show high source completeness (around 90 per cent at maximum) while the darker (lower, right) shows comparatively low completeness (less than 50 per cent). The sudden changes seen at 50 per cent completeness limit denote where the photometric band used for selection changes (at redshifts 3.5 and 4.5). The CFHT-D1 simulation does not extend above redshift 4.5 as for $z\simeq5$ it uses the same imaging as the HSC data from XMM-DEEP.}
\label{fig:Comp}
\end{figure*}

\subsection{Assessing Contamination}

We next examine the contamination rate of low-redshift sources entering the redshift range of interest, which is found to be very low. Across the full sample of $\sim 58,000$ spectroscopically confirmed sources in our three primary fields, we find a total of 12, 4, 1 low-redshift interlopers enter our final sample of galaxies in each of our $2.75<z<3.5$, $3.5<z<4.5$, $4.5<z<5.2$ bins. The spectroscopic sample makes up approximately 2.5 to 4 per cent of the total number of sources detected to $5\sigma$ in each of the three main fields used. However, there is a bias towards spectroscopic sources being more numerous at brighter luminosities. The majority of the sources used to measure the UV LF's in this study occupy the apparent magnitude range of $23.5<m<26.5$. Within this luminosity range, only 1361 sources, or 0.08 per cent of the 1.65 million sources at this luminosity range, have spectroscopic redshifts across our three fields. Under the simple assumption that the contamination rate measured in the spectroscopic sample in this luminosity range continues for the full photometric sample, the total number of contaminants will be approximately 1000 times that found in the spectroscopic sample. This amounts to 6, 9 and 8 per cent contamination rate of the final redshift bins. As the overall contamination is found to be less than 10 per cent and of similar scale to other systematic errors (e.g. cosmic variance), we do not consider any major contamination corrections to the UV LF.


\section{Methods}\label{sec:method}

Armed with our robust samples, we proceed to measure the UV LF, taking into account the completeness of our sample and the effects of cosmic variance.

\subsection{Completeness Simulation}\label{sec:completeness}

The incompleteness of source detection can be described primarily through two effects. The first of these is the probability that an object is lost due to partial or total blending with a secondary source in the image. The second is the dependence on $M_{\rm UV}$, redshift and location within the imaging data on whether a source meets the magnitude cut corresponding to the average $5\sigma$ depth used for selecting a confident sample.

The impact of the first effect can be assessed by examining the segmentation map generated by {\sc SExtractor} for each image. The ratio between the number of unoccupied sky pixels and the total number of unmasked pixels can provide the percentage chance that a randomly positioned source would overlap with another. In this situation, we apply a simple assumption that a highly blended source is one where the centre of the source lies on a pixel already occupied by another, leading to cases where the source is upwards of 50 per cent blended. Such a source is regarded as irretrievable and is assumed to not fall within our source selection. To correct for the second effect, we conduct a simulation where 10.4 million fake sources are injected into the images used for sample selection. Sources are inserted such that there are 200,000 for every 0.25 bin in redshift in each sub-field. To avoid making the images artificially over-dense, we insert only 4000 objects at a time into the images before reprocessing. These sources have an assumed spatial profile that is described by a S\`{e}rsic index of $n=1$ \citep{1963BAAA....6...41S,Conselice2014} and have an intrinsic ultraviolet luminosity that is drawn between $-23<M_{\rm UV}<-18$ from the redshift evolution of the UV LF derived in \citet{Bouwens2015}. This ensures that we account for Eddington bias, where the larger number of faint sources have a chance of being scattered above the $5\sigma$ detection limit. Simulated sources have a half-light radius that is sampled from the results by \citet{Huang2013} for $z=4$ and 5, with the trends found extrapolated to $z\simeq3$. The resultant luminosity profile is then convolved with the PSF model for the respective band and field combination. Our observations however, are seeing-dominated ($\approx 0.8\text{--}0.9\arcsec$) and so the assumptions regarding the intrinsic light profile make negligible difference to our results.

We restrict the simulation from placing galaxies with a central coordinate that is occupied by another source. This prevents double counting the first effect and allows instances for partial ($<50\%$) blending to occur. To add in the probability of total blending to the final completeness estimation, the recovery rate as a function of redshift and UV luminosity is scaled by the ratio of previously unoccupied pixels and the total number of unmasked pixels. The derived completeness curves for each region are presented in Fig.~\ref{fig:Comp}. To minimise the complexity of our selection functions, subfields are only used in the measurement in the final LF if they are considered complete ($>50$ per cent) across the full bin width considered (e.g. $2.75<z<3.5$ for $z\sim3$). In addition, to minimise cosmic variance we require that at least two sub-fields be 50 per cent complete when calculating our final measurement of the UV LF. The implementation of the 50 per cent completeness requirement effectively increases the SNR cuts to the final sample. We find the absolute magnitudes brightwards of this completeness limit correspond to around $7.5\sigma$ detections in each of our sub-fields. The peak completeness of 80-85 per cent in each of our fields matches the recovery rate of the spectroscopic sample.

The authors note that this simulation considers only the incompleteness of successfully identifying a source within the imaging used for detection. There are studies that go on to simulate galaxy colours and reproduce the full sample selection procedure \citep[see e.g.][]{Bouwens2015,Finkelstein2015,Bowler2015,Stevans2018,Ono2017,Bowler2020,Harikane2021}. From experience gathered from the simulations run in \citet{Bowler2015,Bowler2020} on the same data used in this study, we have found the completeness of sources that are detected at high-significance in the data (e.g. the sources at $L \gg L^*$ that lead to the main conclusions in this paper) have completeness values close to unity.

\subsection{Measuring the UV LF with the $1/V_{max}$ method}

We use the 1/$V_{\rm max}$ method \citep{Schmidt1968,rowanrobinson1968} to measure the UV LF of our sample. We calculate the maximum observable redshift $z_{\text{det}}$ by iteratively shifting the best-fit SED of each source in small steps of $\delta z = 0.01$ and convolving this with the selection filter (the $r$,$i$,$z$-bands) until the galaxy falls below the $5\sigma$ detection threshold for the corresponding band and field it is located in. A maximum volume in which the object could have been detected ($V_{\rm max}$) is thus the co-moving volume contained within the range $z_{\text{min}} < z < z_{\text{max}}$, where $z_{\text{min}}$ is the lower boundary of the redshift bin and $z_{\text{max}}$ is either the maximum bound of the redshift bin or $z_{\text{det}}$ if it is found to be lower. From this, the rest frame UV LF ($\Phi(M)$) is calculated using:

\begin{equation}\label{eqn:lf}
\Phi(M) d \log(M) = \frac{1}{\Delta M } \sum_i^N \frac{1}{C_{i,f} V_{\rm max,i}} ,
\end{equation}
where $\Delta M$ is the width of the magnitude bins and $C_{i,f}$ is the completeness correction for a galaxy $i$ depending on its location within a sub-field, $f$. 

The measured uncertainty of the LF is given by:

\begin{equation}
\delta \Phi(M) = \frac{1}{\Delta M} \sqrt{\sum_i^N \left(\frac{1}{V_{\rm max,i}}\right)^2} .
\end{equation}

To balance number statistics with resolution in luminosity, we utilise four bin widths of $\Delta M = 0.25,0.5,0.75,1.0$. Bin widths of $1.0$ are used at $M_{\rm UV} < -23.5$ for the $z=5$ UV LF, while widths of $0.75$ are used for $z=4$ and $z=3$ due to their greater number statistics. Widths of 0.5 are used in the intermediate regime of $-23.5 < M_{\rm UV} < -23$ and widths of 0.25 are used for $M_{\rm UV} > -23$. The absolute UV magnitude ($M_{\rm UV}$) is calculated by convolving a 100\angstrom\, top-hat filter, centred at \SI{1500}{\angstrom} within the rest frame, on the best-fitting SED of each object. Bins fainter than our imposed completeness limits contain large number of candidates due to the shape of the UV LF. This means the final sample used to measure the LFs in this study contain fewer sources than the original selection procedure. The final source counts used to derive the complete UV LF are 96,894 sources for $z\simeq3$, 38,655 sources for $z\simeq4$ and 7,571 sources for $z\simeq5$. The mean redshifts of the final samples are $\overline{z}=3.1$, $\overline{z}=4.0$ and $\overline{z}=4.8$ respectively.

\subsubsection{Cosmic variance}\label{sec:CV}

As the UV LF is measured using a finite volume of the Universe and three sight-lines, it can be susceptible to biases resulting from the large-scale structure of the Universe. Finite sight-lines can cause non-representative conclusions to be drawn and such an effect is commonly referred to as `cosmic variance'. We follow the same procedure implemented in \citet{Adams2020}, which uses the online cosmic variance calculator produced in \citet{Trenti2008}\footnote{https://www.ph.unimelb.edu.au/~mtrenti/cvc/CosmicVariance.html} to compute the additional error due to cosmivc variance. The total survey area used to measure the LF decreases towards fainter luminosities as the shallower sub-fields fall below 50 per cent completeness. This means our cosmic variance estimations vary across the full luminosity range probed. At brighter intrinsic luminosities, the error budget is dominated by low number statistics, while cosmic variance dominates at fainter luminosities. To be conservative, we round up the value obtained from the calculator to the nearest whole percentage point, ranging from 3 per cent for bright $z\simeq3$ luminosity bins to 9 per cent for the faintest $z\simeq5$ luminosity bins, the resultant error is then added in quadrature to our LF uncertainty resultant from counting statistics in Equation 2. 

\section{Results}\label{sec:LFs}

In this section, we present the raw (binned) results of our measurements of the ultraviolet luminosity function and the results of our subsequent fitting procedures.

\subsection{The Binned UV LF at $z=3-5$}

\begin{figure*}
    \centering

    \includegraphics[width=0.85\textwidth]{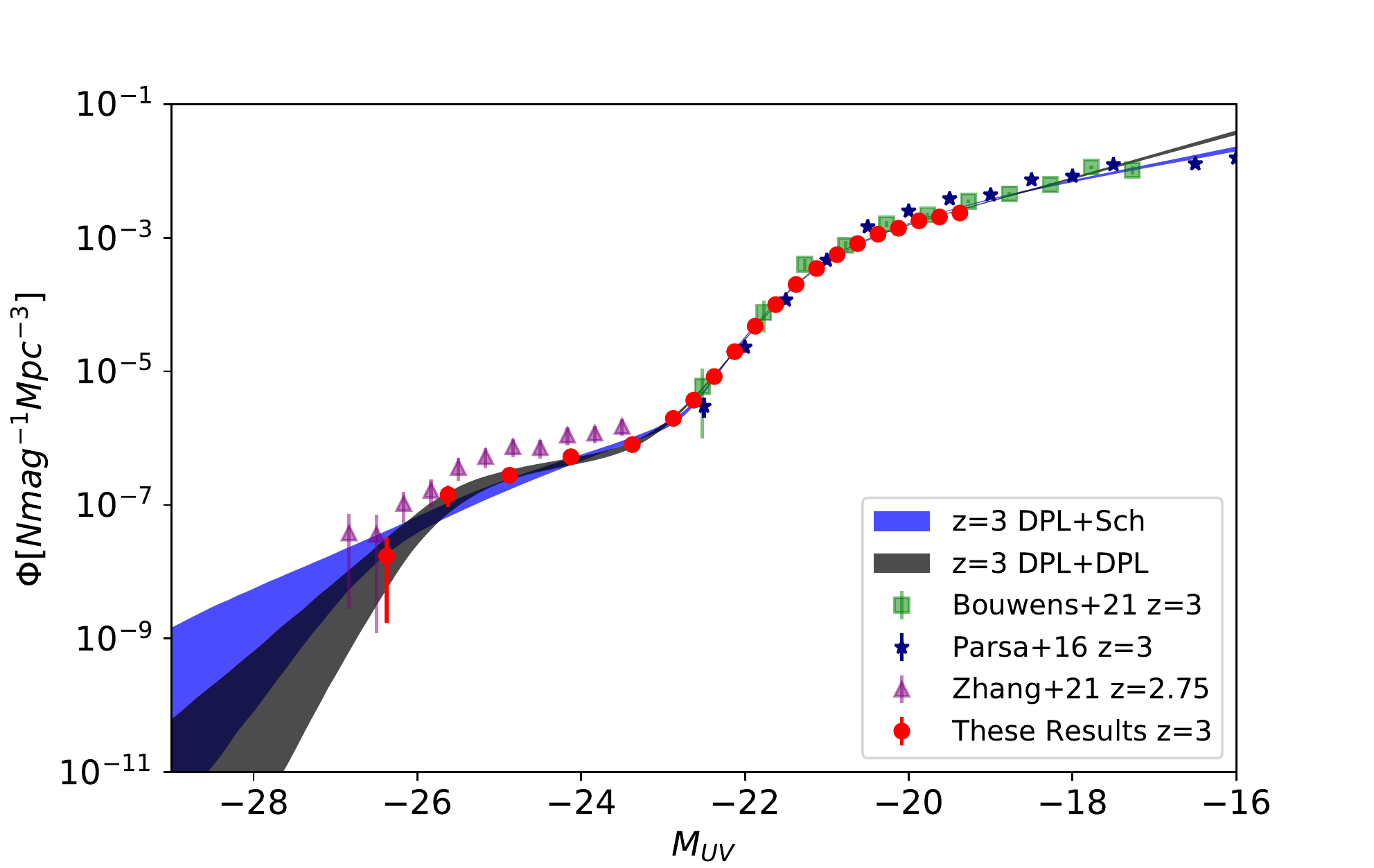}
    \caption{The measured rest-frame ultraviolet luminosity function at $2.75<z<3.5$ as measured in this work and in comparison to previous studies. The red data points are those measured by this study across the COSMOS, XMM-LSS and CDFS fields. The purple triangles are the AGN UV LF as measured by \citet{Zhang2021}, dark blue stars are from \citet{Parsa2016} and the green squares are by \citet{Bouwens2021}. The shading displays the results from the MCMC model fitting procedure. The DPL+Sch functional form is in blue with its $1\sigma$ uncertainty indicated by the width of the shaded line, the DPL+DPL functional form is displayed in grey.}
    \label{fig:Z3F}
\end{figure*}

\begin{figure*}
    \centering

    \includegraphics[width=0.85\textwidth]{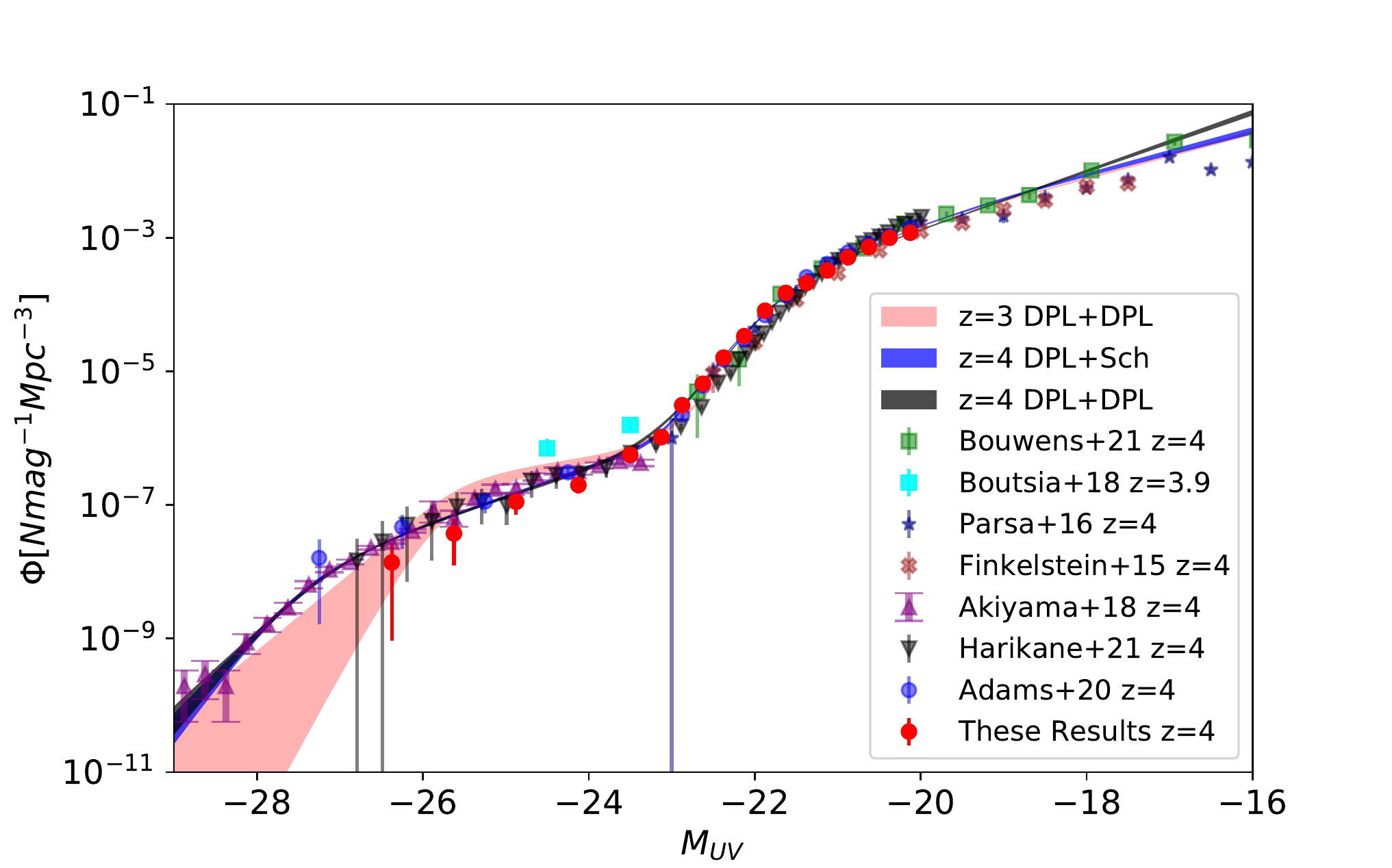}
    \caption{The measured ultraviolet luminosity function at $3.5<z<4.5$ as measured in this work and in a selection of others. The red data points are those measured by this study across the COSMOS, XMM-LSS and CDFS fields. The purple triangles are the AGN UV LF as measured by \citet{Akiyama2018}, the blue circles are past results from \citet{Adams2020}, the black downwards triangles are results from \citet{Harikane2021}, dark blue stars are from \citet{Parsa2016}, brown crosses are from \citet{Finkelstein2015} and the green squares are by \citet{Bouwens2021}. The shading displays the results from the MCMC model-fitting procedure. The DPL+Sch functional form is in blue with its $1\sigma$ uncertainty indicated by the width of the shaded line, the DPL+DPL functional form is displayed in grey. The red line shows the result of the best-fit $z=3$ DPL+DPL.}
    \label{fig:Z4F}
\end{figure*}

Following the processes outlined in Section \ref{sec:method}, we obtain the UV luminosity functions presented in Fig. \ref{fig:Z3F}, Fig. \ref{fig:Z4F} and Fig. \ref{fig:Z5F} for the $z=3$, $z=4$, $z=5$ bins respectively. The raw data points for each UV LF are provided in Table~\ref{Tab:Points}. We display and use only the bins estimated to be greater than 50 per cent complete from at least two different sub-fields in our completeness simulation.

For the $z=3$ UV LF, we are capable of measuring the number density of sources over a very wide range of intrinsic luminosities, spanning $-26.75<M_{\rm UV}<-19.25$ and 6 orders of magnitude in number density. It encompasses the characteristic `knee' of both the AGN and galaxy luminosity functions in addition to sampling the transition between AGN and galaxy dominated number counts. The measured number density of AGN is found to be slightly lower than the UV LF of AGN measured by \citet{Zhang2021}. However, this can be explained by the different redshift binning used between our two studies, with \citet{Zhang2021} using a bin of $2.0<z<3.5$, resulting in a lower mean redshift. For the bright galaxies $-23<M_{\rm UV}<-21$, we have very strong agreement with two primary observations of this regime by \citet{Parsa2016} and \citet{Bouwens2021}. Our large area enables us to more finely bin across this luminosity regime, providing 2-3 times the resolution while maintaining smaller observational errors. At the faint end, our galaxy LF is slightly lower than found by \citet{Parsa2016} and more closely matches the observations from \citet{Bouwens2021}, this could again be due to our slightly higher mean redshift compared to these two studies, an aspect we explore further in Section \ref{sec:fits}.

The measured $z=4$ UV LF spans a slightly smaller luminosity interval than the $z=3$ UV LF, with a final range of $-26.75<M_{\rm UV}<-20$. AGN measurements broadly agree with those from \citet{Akiyama2018} and the transition between AGN and galaxy-dominated number counts matches recent observations presented in \citet{Adams2020} and \citet{Harikane2021}. Our observations are found to disagree with the findings from \citet{Boutsia2018}, whom find an excess of faint-end AGN with $M_{\rm UV} \sim-23.5$. We find that the observation from \citet{Boutsia2018} that is uncorrected for completeness lies much closer to observations of this study and \citet{Harikane2021}, indicating that the completeness correction employed within that study may be too strong. The number density of bright galaxies ($-23<M_{\rm UV}<-21$) are found to sit at the upper end of the range previously found by past attempts to measure the UV LF, agreeing more with \citet{Parsa2016} and \citet{Bouwens2021} than \citet{Finkelstein2015} and \citet{Harikane2021}. Agreement with previous observations continues through the lower intrinsic luminosities, with our measurements towards the centre of the distribution of number densities found in \citet{Finkelstein2015}, \citet{Parsa2016} and \citet{Bouwens2021}.

Finally, our $z=5$ UV LF covers a luminosity range of $-25.5<M_{\rm UV}<-20.5$. The use of 10\ds\, of sky is insufficient to probe the AGN LF to higher luminosity due to the significantly lower number densities that are present at this epoch when compared to $z=4$ and $z=3$. The few ultra-luminous sources that we do measure at $M_{\rm UV}<-24$ match the measurements of the AGN LF from \citet{Niida2020} and \citet{Harikane2021}, which use substantially larger areas in their studies but lack the ancillary near-infrared information used in this study. The number density and shape of the transition between AGN and galaxy-dominated number counts match those measured by \citet{Harikane2021}. As with our other luminosity functions, the galaxy component is also in strong agreement with past observations from \citet{Bouwens2021} and slightly higher than those of \citet{Finkelstein2015}. The study by \citet{Finkelstein2015} attributes their lower number density of faint galaxies to the use of \emph{Spitzer} data to remove contaminants in the form of low-redshift galaxies and Milky Way brown dwarfs. With the use of ground-based near-infrared data, we find that brown dwarfs should be sufficiently discarded up to redshifts of $z=5.2$, beyond which optical colours of galaxies become too similar to dwarf stars and the VISTA bands do not probe to red enough wavelengths to capture the expected turn-over in the spectrum of a cold, dwarf star.

\subsection{Fitting the Luminosity Function}\label{sec:fits}

With the measurement of the UV LF completed, we proceed to fit a number of parametric models. Since we do not differentiate AGN and galaxies, we elect to simultaneously fit both the AGN and galaxy UV LFs. The two primary models we use consist of a DPL for the AGN and either a Schechter or DPL for the galaxy population. We do this to assess which of the DPL or Schechter functional forms better fit the galaxy population. We combine our LF results with those of other studies that probe luminosity regimes beyond what is possible with the dataset used here. Data used to populate the very faint end of the galaxy population is obtained from \citet{Bouwens2021}. This data is included in order to provide tighter constraints on the faint-end slope ($\alpha$) of the galaxy LF. We find that the precision of the measured value of $\alpha$ is improved by a factor of two with the inclusion of the \citet{Bouwens2021} data as opposed to fitting our measured LF alone. We only use the \citet{Bouwens2021} data that is fainter than the completeness limit for each of our UV LFs, providing total coverage as faint as $ M_{\rm UV} = -16$. 

For the AGN UV LF, our $z=3$ measurements probe beyond the `knee', which enables us to fit the whole AGN LF with the use of our data points alone. However, for $z=4$ and $z=5$, our lack of survey volume prevents us from constraining the bright-end slope and knee location. We subsequently employ the use of recent results from wide-field Subaru/HyperSuprimeCam studies which cover similar redshift bins to those used in this study. These are \citet{Akiyama2018} and \citet{Niida2020} for the $z\simeq4,5$ redshift bins respectively. We only utilise data points with $M_{\rm UV} < -23.5$ from the AGN studies. This cut is used for two primary reasons to ensure our UV LFs include all sources with rest-frame ultraviolet emission. Firstly, the study conducted in \citet{Akiyama2018} uses a morphological selection procedure for their AGN sample. This favourably selects strongly Type-I AGN which have point-source morphology. In \citet{Bowler2021}, we show that for magnitudes fainter than $M_{\rm UV} > -23$, such a morphological selection underestimates the total number sources with AGN. Secondly, \citet{Adams2020} has shown that galaxies begin to dominate the total number counts of the sources between $-23.5 < M_{\rm UV} < -23$ at these redshifts.

Because the data points obtained from \citet{Bouwens2021} at the far faint end have slightly different mean redshifts to the UV LFs derived in this study, we make a simple modification that uses the evolutionary model of the UV LF derived in \citet{Bouwens2021} to shift these data points to match the mean redshift of our observations. This is done in order to minimise any potential discontinuity from affecting our results. We find that fitting to these modified points lead to higher quality fits with smaller $\chi^2$ values, indicating that discontinuities between the two data sets have been reduced.

Our fitting procedure uses a Markov-Chain Monte Carlo (MCMC) that is implemented using {\tt emcee} \citep{emcee}. The MCMC has 500 walkers which each burn in for 100,000 steps before mapping the posterior distribution for a further 20,000 steps. The walkers are initially distributed uniformly over a wide parameter space and priors for each parameter are set to be wide and uniform. The results of this procedure are presented in Table \ref{tab:Result3} for the three redshift bins. In the following subsections, we discuss our findings from this fitting procedure.

\subsubsection{Results of the $z\sim3$ UV LF fits}

For the $z=3$ UV LF, our DPL+Sch model fits obtain a value of $M^* = -20.59^{+0.03}_{-0.03}$ with a faint-end slope of $\alpha=-1.52^{+0.03}_{-0.03}$ for the galaxy UV LF. The DPL+DPL model fits, we find best-fit values of $M^*=-21.18^{+0.03}_{-0.03}$, $\alpha=-1.85^{+0.02}_{-0.02}$ and  $\beta=-4.95^{+0.08}_{-0.09}$, resulting in greater number of both the most luminous and the faintest galaxies compared to the Schechter parameterisation. The bright end of the AGN LF is poorly constrained, but the faint end is measured to be $\alpha_{\rm AGN}=-2.10^{+0.22}_{-0.14}$, when using the Schechter parameterisation for the galaxies, and $\alpha_{\rm AGN}=-1.37^{+0.23}_{-0.23}$ when using the DPL parameterisation for the galaxies, showing that more UV-faint AGN are required if the galaxy UV LF exponentially falls off. We find that the reduced $\chi^2$ ($\chi^2_{red}$) values at the peak of the posterior are over a factor of two lower when using a DPL to describe the galaxy UV LF as opposed to using a Schechter function. Detailed examination shows that our observations at $M_{\rm UV}=-19.125$ and the two faintest points from \citet{Bouwens2021} at $M_{\rm UV}\sim-17$ are the largest contributors to the total $\chi^2$. These results show a discontinuity is still present between the two data sets, even after our adjustments. This could be the consequence of a different selection function within the redshift bin or an underestimation of the total cosmic variance impacting either study.

We find that the DPL+DPL results produce a faint-end slope for the AGN UV LF that agrees with the findings presented in \citet{Zhang2021} ($\alpha_{\rm AGN}=-1.26^{+0.10}_{-0.08}$). The work in \citet{Zhang2021} fits for just the Type-I (broad line) AGN population. However, they note that Type-II AGN (narrow line) become dominant in the AGN LF at $M_{\rm UV}>-22$, a regime where our LF is dominated by star-forming galaxies. It is thus no surprise that our fits replicate the LF slope of Type-I AGN. Compared to the UV LFs derived in \citet{Harikane2021}, the differences between our measurements are primarily driven by the higher number density of bright, star-forming galaxies found at $M_{\rm UV}<M^*$ in \citet{Moutard2020}, the source of the bright-end galaxy LF data points used in \citet{Harikane2021}. The higher number density of galaxies at $M_{\rm UV}\sim-23$, lead to a transition between the galaxy and AGN LF that is less sharp than found by our observations. This results in \citet{Harikane2021} finding a value of $\alpha_{AGN}\sim-1.59$ when using a DPL to describe the galaxy population, steeper than found here and by \citet{Zhang2021}. The cause of these differing results across the transition between the galaxy and AGN LFs can be attributed to the way in which the absolute magnitudes are measured. In this study, we place a 100\AA\, top hat across 1500\AA\, in the rest frame using the best-fit SED, while in \citet{Moutard2020} the closest observer frame photometric band is used, which have more complicated shapes and are vulnerable to being affected by sources with steep ultraviolet spectral slopes. This results in an additional source of scatter in the $M_{\rm UV}$ measurement, leading to an artificial increase in the number of apparently luminous galaxies brightward of the knee \citep[see appendix of][for an explanation of the differences in $M_{\rm UV}$ measurements and the shallower galaxy-AGN transition]{Adams2020}.

\begin{figure*}
    \centering

    \includegraphics[width=0.85\textwidth]{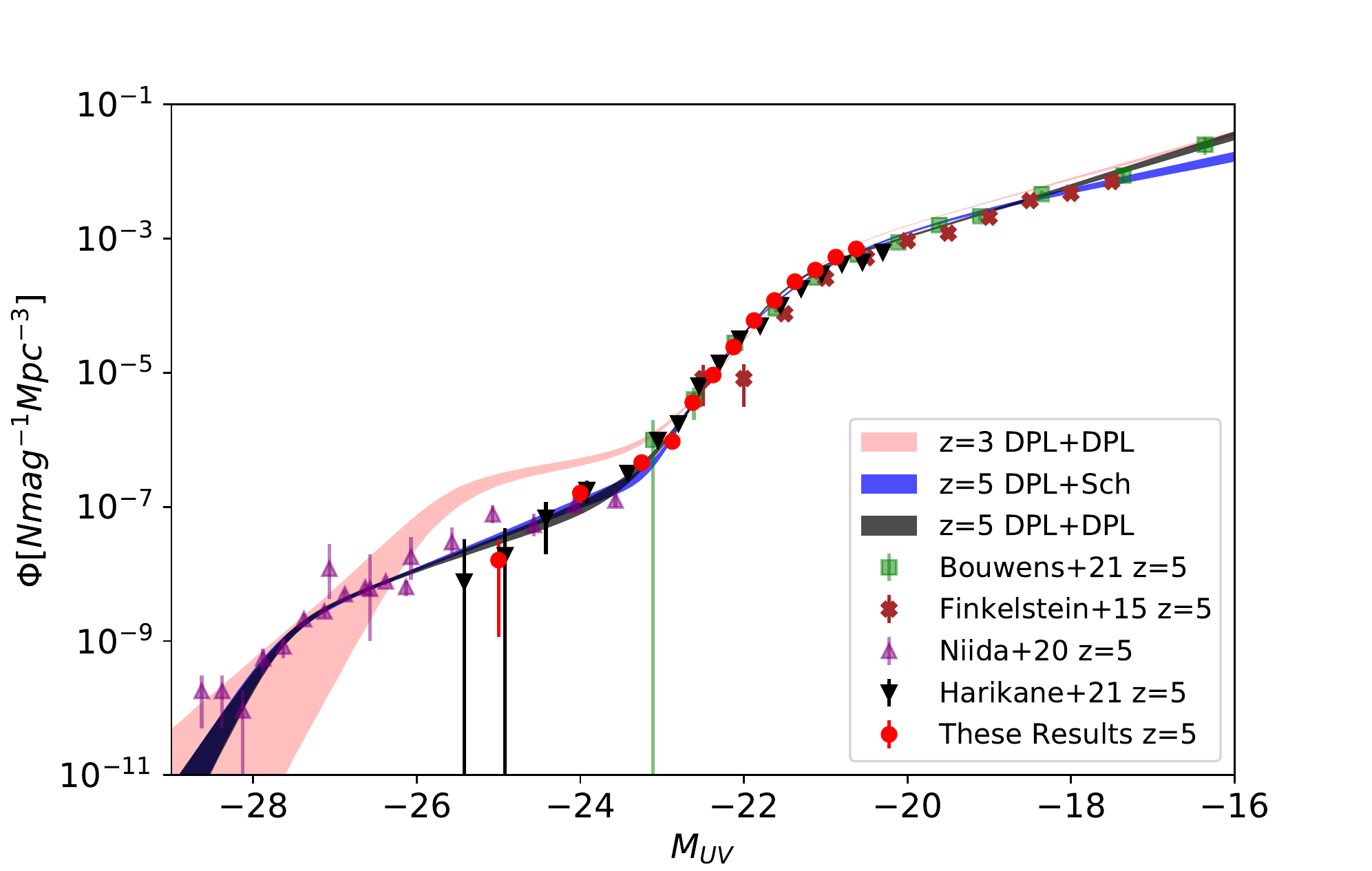}
    \caption{The measured ultraviolet luminosity function at $4.5<z<5.2$ as measured in this work and in a selection of others. The red data points are those measured by this study across the COSMOS, XMM-LSS and CDFS fields. The purple triangles are the AGN UV LF as measured by \citet{Niida2020}, the black downwards triangles are results from \citet{Harikane2021}, brown crosses are from \citet{Finkelstein2015} and the green squares are by \citet{Bouwens2021}. The shading displays the results from the MCMC model-fitting procedure. The DPL+Sch functional form is in blue with its $1\sigma$ uncertainty indicated by the width of the shaded line, the DPL+DPL functional form is displayed in grey. The red line shows the result of the best-fit $z=3$ DPL+DPL.}
    \label{fig:Z5F}
\end{figure*}

When the galaxy population is instead modelled by a Schechter function, the number density of faint AGN is estimated to be much higher in order to account for the more rapid decline in galaxies. This results in a faint-end slope that is much steeper than in the DPL case, with $\alpha_{\rm AGN}=-2.10^{+0.22}_{-0.14}$. This diversion from the \citet{Zhang2021} AGN LF results, derived from spectroscopic measurements, and the lower quality of fit (higher $\chi^2_{\rm red}$) provides evidence that the galaxy UV LF is better described by a double power law at $z\sim3$.

\begin{table*}
\centering
\caption{The results of the MCMC fitting applied to the total UV LF in each of our redshift bins. The fit uses data points from \citet{Bouwens2021} fainter than probed by our measurements. Additionally, the $z\simeq4$ and $z\simeq5$ UV LFs use the measured AGN LF from \citet{Akiyama2018} and \citet{Niida2020} respectively.
The first column lists the fitting parameterisation used. Columns 2--5 are the best fit DPL parameters for the AGN LF. Columns 6--9 show the best fit LBG parameters, either for a Schechter function or DPL.
The final columns show the $\chi^2$ and reduced $\chi^2$ of the fit.
We modify the data points from \citet{Bouwens2021} using their parameterised evolution of the LF parameters to shift the mean redshift to match the mean redshift of our data.}\label{tab:Result3}
\begin{tabular}{l|llllllll|ll}
\hline
Function  & $\textrm{log}_{10}(\Phi_{\text{AGN}})$ & $M^*_{\text{AGN}}$ & $\alpha_{\text{AGN}}$ & $\beta_{\text{AGN}}$ & $\textrm{log}_{10}(\Phi)$ & $M^*$ & $\alpha$ & $\beta$ & $\chi^2$ & $\chi_{\rm red}^2$ \\
  & $\textrm{mag}^{-1} \textrm{Mpc}^{-3}$ & mag &  &  & $\textrm{mag}^{-1} \textrm{Mpc}^{-3}$ & mag &  &  &  &  \\\hline
$z=3.1$        &                      &                           &       &          &         &          & &  & &                     \\                    
DPL+Sch            & $-7.46^{+0.65}_{-1.08}$   & $-26.67^{+0.99}_{-2.12}$ & $-2.10^{+0.22}_{-0.14}$ &  $-4.10^{+1.44}_{-1.30}$   &  $-2.63^{+0.03}_{-0.03}$ & $-20.59^{+0.03}_{-0.03}$   & $-1.52^{+0.03}_{-0.03}$      &  --        &  70.83       &    4.17                         \\
DPL+DPL              & $-6.57^{+0.21}_{-0.24}$   & $-25.57^{+0.47}_{-0.37}$ & $-1.37^{+0.23}_{-0.23}$ &   $-4.91^{+1.37}_{-1.89}$   &  $-3.20^{+0.03}_{-0.03}$ & $-21.18^{+0.03}_{-0.03}$   & $-1.85^{+0.02}_{-0.02}$      &  $-4.95^{+0.08}_{-0.09}$        &  28.75       & 1.80                             \\ \hline
$z=4.0$        & &                           &       &          &         &          & &  & &                     \\
DPL+Sch               & $-7.90^{+0.15}_{-0.11}$   & $-27.34^{+0.20}_{-0.14}$ & $-2.10^{+0.06}_{-0.05}$ &  $-4.64^{+0.44}_{-0.53}$                     &  $-3.00^{+0.03}_{-0.03}$ & $-21.11^{+0.04}_{-0.04}$   & $-1.80^{+0.03}_{-0.03}$      &  ---        &  74.32        &  2.06                  \\
DPL+DPL               & $-7.77^{+0.21}_{-0.15}$   & $-27.18^{+0.29}_{-0.19}$ & $-2.02^{+0.10}_{-0.07}$ &  $-4.34^{+0.43}_{-0.51}$                     &  $-3.62^{+0.04}_{-0.04}$ & $-21.68^{+0.05}_{-0.04}$   & $-2.10^{+0.03}_{-0.03}$      &  $-5.29^{+0.14}_{-0.15}$         &  84.44       &  2.41                \\ \hline
$z=4.8$        &                      &                           &       &          &         &          & &  & &                     \\  
DPL+Sch                & $-8.74^{+0.12}_{-0.11}$   & $-27.65^{+0.13}_{-0.13}$ & $-2.23^{+0.10}_{-0.08}$ &  $-6.54^{+1.05}_{-0.95}$   &  $-2.88^{+0.04}_{-0.04}$ & $-20.86^{+0.05}_{-0.05}$   & $-1.57^{+0.06}_{-0.05}$      &  --        &  43.86       &    1.46                        \\
DPL+DPL                 & $-8.65^{+0.13}_{-0.12}$   & $-27.58^{+0.14}_{-0.13}$ & $-2.12^{+0.12}_{-0.10}$ &   $-6.31^{+1.04}_{-1.04}$   &  $-3.57^{+0.05}_{-0.05}$ & $-21.60^{+0.05}_{-0.05}$   & $-1.94^{+0.04}_{-0.04}$      &  $-5.73^{+0.18}_{-0.19}$        & 24.04        &  0.83                         \\ \hline

\end{tabular}
\end{table*}

\subsubsection{Results from the $z\sim4$ UV LF fits}

For the $z=4$ UV LF, our DPL+Sch model fits obtain a value of $M^* = -21.11^{+0.04}_{-0.04}$ with a faint-end slope of $\alpha=-1.80^{+0.03}_{-0.03}$ for the galaxy UV LF. The DPL+DPL model fits, we find best-fit values of $M^*=-21.68^{+0.05}_{-0.04}$, $\alpha=-2.10^{+0.03}_{-0.03}$ and  $\beta=-5.29^{+0.14}_{-0.15}$. As with the $z=3$ UV LF, the DPL parameterisation predicts more galaxies at either extreme of the $z=4$ galaxy UV LF compared to using a Schechter parameterisation. A key finding is that there are fewer galaxies with $M_{\rm UV}<-22$ at $z\simeq3$ compared to $z\simeq4$, which we discuss in more detail in Section \ref{sec:lfevo}.

For the AGN UV LF, we find that both parameterisations of the galaxy LF that we fit provide consistent measurements for the DPL parameters used to describe the AGN UV LF. Our fitting procedure produces faint-end slopes that are quite steep with $\alpha_{AGN}<-2$. This result contrasts with the findings of the original \citet{Akiyama2018} study ($\alpha_{AGN}\sim1.3$), but broadly agrees with other studies that have sought to simultaneously fit for both AGN and galaxies \citep{Stevans2018,Adams2020,Harikane2021}. This disagreement with \citet{Akiyama2018} can be attributed to the morphological selection that they employ, which \citet{Bowler2021} shows could lead to an underestimation of the number of sources with AGN signatures at $M_{\rm UV} > -23$ \citep[and one of the reasons we employed cuts on which data points from ][were used]{Akiyama2018} . The differing faint-end slope for the AGN LFs can subsequently be attributed to the handling of sources that blur the boundary between the definitions of star-forming and AGN-dominated UV emission when spectroscopic data is lacking.

While the $z\sim3$ UV LF has a clear distinction between the quality of the fits of the Schechter and DPL functional forms for the galaxy population, at $z\sim4$ the different models are lower in the quality of fit and more comparable in their significance. In contrast to the results from $z=3$ (and $z=5$), the $z=4$ UV LF shows that the Schechter function is slightly more favourable than the DPL functional form. As with the $z=3$ UV LF, we find that our modifications to the \citet{Bouwens2021} points reduces the discontinuity at the faint-end as well as the overall $\chi^2$. However, we find that the data points straddling this transition and the far faint-end are still responsible for around half of the total $\chi^2$ contributions. For $M_{\rm UV} < -21$, the typical $\chi^2$ contributions are around 1, indicating that the rest of the UV LF is fit well.

The far faint end of the $z=4$ UV LF has a greater tension between past observational studies than the other redshift bins considered in this study. With the results from \citet{Finkelstein2015} and \citet{Parsa2016} clearly offset to lower number density than \citet{Bouwens2021}. To assess what impact the choice of study used to populate the faint-end has on our results, we repeat our MCMC fitting procedure using the \citet{Finkelstein2015} data instead of \citet{Bouwens2021}. The results of this procedure are displayed in Appendix \ref{sec:AppB}. We find that the AGN UV LF is unaffected. For the galaxy UV LF, we find that the Schechter parameterisation experiences significant shifts towards higher values of $\Phi^*$, fainter characteristic magnitudes and flatter faint-end slopes. Compared to the results obtained with \citet{Bouwens2021}, the fits using \citet{Finkelstein2015} data find $\delta\Phi=0.13$ dex, $\delta M^*=0.19$ magnitudes and the faint-end slope flattens from $\alpha=1.80$ to $\alpha=1.49$. For the DPL+DPL fits, we find that the galaxy faint-end slope is also flatter (with $\delta\alpha=0.25$) compared to \citet{Bouwens2021}. The remaining parameters shift to a similar degree as the Schechter parameterisation, with $\delta\Phi=0.19$ dex, $\delta M^*=0.19$ magnitudes. The bright-end slope changes by a lower significance, with the shift around $2\sigma$. This shows that the bright-end slope is largely constrained by our new measurements of the galaxy bright-end and ‘knee’. Overall, the fits with the \citet{Finkelstein2015} data points at $z=4$ provide a smoother evolution between our other fits at $z=3$ and $z=5$, but it is clear that systematics at the faint-end still remain at $z = 4$, leading to uncertainties in the faint-end. With the recent releases of the first data from the \emph{James Webb Space Telescope} (JWST) we will soon see new capabilities for improved photometric redshifts, spectroscopic completeness and more at $M_{\rm UV} > -20$, which will help alleviate this issue.

\begin{table*}
\caption{The evolution parameter $k$ for the faint end of the AGN LF as calculated with the primary fits to the combined AGN and galaxy UV LF. The first column indicates the absolute luminosity at which the $k$ value was measured, the following columns then present the measured $k$ value between two neighbouring redshift bins ($3-4$ or $4-5$) and using either a Schechter (Left) or DPL (right) parameterisation for the galaxy LF. Errors calculated by sampling the posteriors generated by the MCMC's used in measuring the UV LF.}\label{tab:EvoParam}
\begin{tabular}{l|ll|ll}
 & Schechter &  & DPL & \\
$M_{\rm UV}$ & $k(z=3-4)$ & $k(z=4-5)$ & $k(z=3-4)$  & $k(z=4-5)$ \\ \hline
-26      & $-0.03\pm0.14$ & $-0.60\pm0.03$ &  $+0.01\pm0.28$                &   $-0.62\pm0.04$               \\
-25      & $-0.12\pm0.09$ & $-0.56\pm0.06$ &   $-0.29\pm0.09$               &  $-0.61\pm0.07$   \\
-24      & $-0.14\pm0.05$ & $-0.51\pm0.09$ &   $-0.11\pm0.06$               &    $-0.56\pm0.10$   \\ \hline

\end{tabular}
\end{table*}

\subsubsection{The $z\sim5$ UV LF fits}

For the $z=5$ UV LF, our DPL+Sch model fits obtain a value of $M^* = -20.86^{+0.05}_{-0.05}$ with a faint-end slope of $\alpha=-1.57^{+0.06}_{-0.05}$ for the galaxy UV LF. The DPL+DPL model fits, we find best-fit values of $M^*=-21.60^{+0.05}_{-0.05}$, $\alpha=-1.94^{+0.04}_{-0.04}$ and  $\beta=-5.73^{+0.18}_{-0.19}$. As with the previous two bins considered, the DPL+DPL parameterisation results in greater numbers of the brightest and faintest galaxies compared to the Schechter parameterisation. 

The Schechter fits to the galaxy LF produce a slightly different set of results compared to those obtained in \citet{Bouwens2021}, with a flatter faint-end slope and fainter $M^*$ luminosity. This is driven by our slightly higher number density at the faint extrema of our observations ($M_{\rm UV} = -20.5$). The DPL parameters that we fit agree well with the fits conducted in \citet{Bowler2020} and \citet{Harikane2021}. We find the transition between the galaxy and AGN LF to be slightly steeper than found by \citet{Harikane2021}, driven by their slightly higher number density at $M_{\rm UV} \sim-23$. This results in a shallower bright-end slope of $\beta=-4.9^{+0.08}_{-0.08}$ in \citet{Harikane2021} compared to the $\beta=-5.73^{+0.18}_{-0.19}$ value obtained in this study. The fitting procedures for our $z=5$ UV LF show a preference for the DPL functional form to describe the galaxy population, with a significantly lower $\chi^2_{\rm red}$ value compared to when a Schechter function is used. However, using the DPL functional form can be considered an overfit to the data, with $\chi^2_{red} < 1$.

For the AGN LF, we find similar results to those found at $z\sim4$. The results of both functional forms provide parameters describing the UV LF of AGN that are consistent at the $\sim1\sigma$ level. The fits produce very steep values for the faint-end slope (with $\alpha_{AGN}<-2$), indicating that there are high numbers of AGN at $M_{\rm UV}>-23$. The bright-end slope $\beta_{AGN}$ is largely unconstrained due to the very low number densities of AGN brighter than $M_{\rm UV}<-28$ at this time and the limited survey volumes probed by our study and that of \citet{Niida2020}. The steep faint-end slope is in agreement with the fits conducted by \citet{Niida2020}, who find $\alpha_{AGN}=-2.0^{+0.40}_{-0.03}$ when the bright end is left free (as opposed to being fixed to a particular value, $\beta_{\rm AGN}=-2.9$ in their case). Similarly, our best fit parameters are close to the results derived from the combined AGN and galaxy LF fit conducted in \citet{Harikane2021}.

\section{Discussion}\label{sec:discussion}

The previous section discusses the results of our UV LF measurements in a base context of comparing different parameterisations for its modelling and direct comparisons to measurements from other studies. In this section, we expand discussion to how these new measurements of the UV LF impact our understanding of the co-evolution of galaxies and AGN.

\subsection{The evolving faint-end slope of the AGN LF}

As shown in Figure \ref{fig:Z5F} and in our fitting in Table \ref{tab:Result3}, we see a strong evolution in the number density of AGN between $z\simeq5$ and $z\simeq3$. To explore this further, we examine the luminosity-dependent evolution of the faint-end slope of the AGN UV LF by comparing the simple vertical change in number density as a function of absolute luminosity. Such an evolution can be described with an "evolution parameter" ($k$), which has been commonly used in previous studies of AGN at high redshifts \citep{Jiang2016,Wang2019,Yang2019,Niida2020}. This parameter follows the definition $k = \log_{10}(\Phi_{z_1}/\Phi_{z_2})/(z_1-z_2)$, where $z_1$ and $z_2$ are the mean redshifts of two luminosity functions used for comparison. The redshift and luminosity dependence of this evolution parameter provides another quantitative method to measure the change in the faint-end slope of the AGN LF. The measurement of $k$ and its subsequent error margin is calculated by randomly sampling the posterior outputs of the MCMC procedure for each redshift bin and examining the spread of $k$ values obtained by comparing neighbouring redshift bins ($z\sim3-4$ and $z\sim4-5$) with each other at three different luminosities ($-26, -25, -24$). The results of this calculation are shown in Table \ref{tab:EvoParam}.

Our measured values of $k$ clearly show that the evolution in number density of UV-faint AGN is much more rapid at higher redshifts; this is further emphasised when examining other studies that have compared the AGN LF between $z\sim5-6$ and find that $k$ is stronger still \citep[$k\sim-0.82$ --- $-0.95$][]{Matsuoka2018c,Niida2020}. The number densities of UV-faint AGN rise by just under two orders of magnitude in the 1Gyr that separates $z\simeq6$ and $z\simeq3$, yet the first order of magnitude growth takes place in only the first $\sim200$Myr of this time frame, providing strong evidence for a very rapid onset of AGN activity following the conclusion of the reionization epoch. This evolution is found to slow as time advances.

\begin{figure*}
    \centering
    \includegraphics[width=1.7\columnwidth]{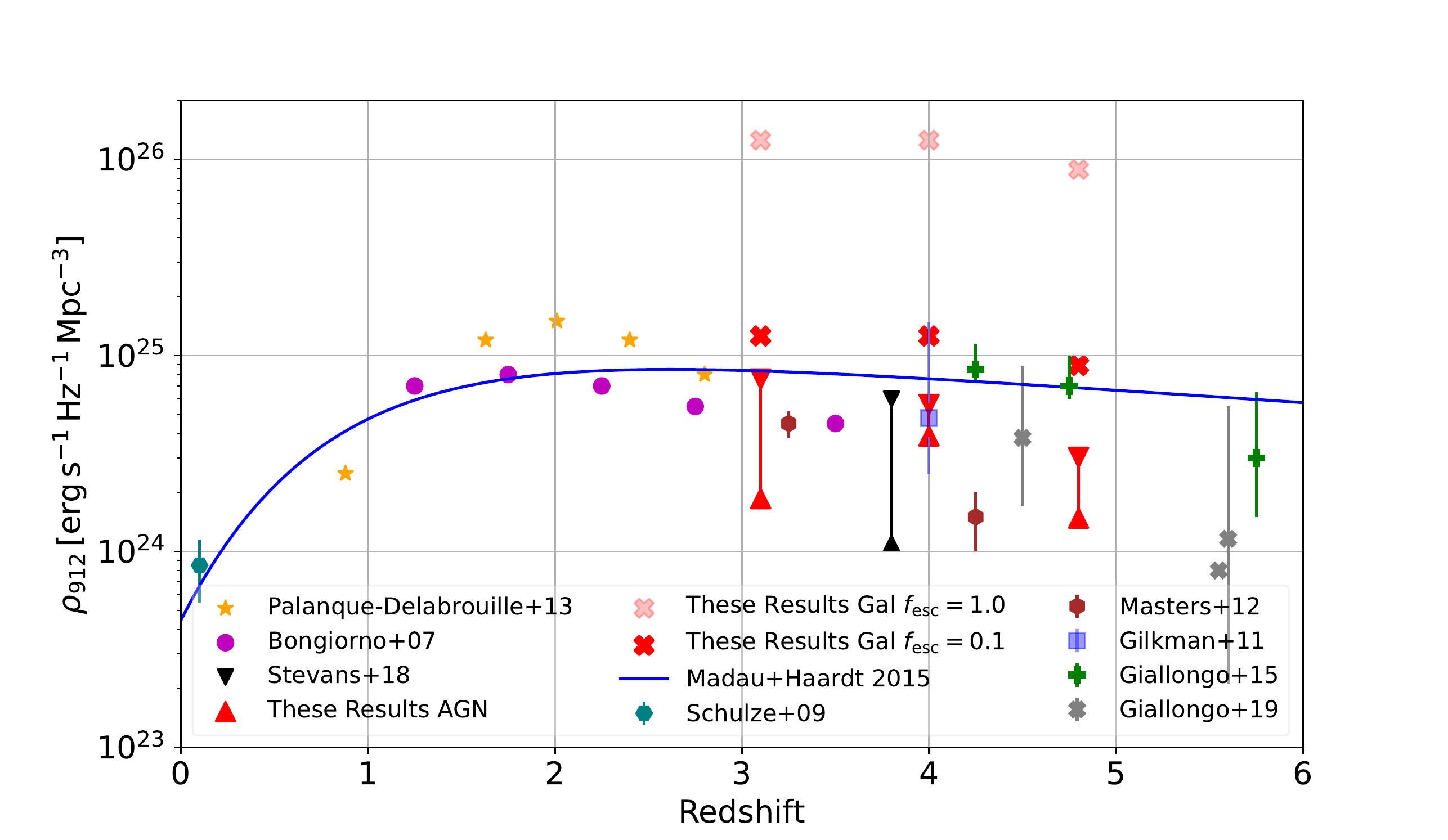}
    \caption{The emissivity of Hydrogen ionizing photons originating from AGN sources. The blue line from \citet{Madau2015} presents a scenario where AGN are capable of driving the reionization of the Universe without the need for contributions from star-forming galaxies. For our data points, the downwards pointing triangles are calculated from the Sch+DPL fits calculated in this study and the upwards pointing triangles are from the DPL+DPL fits. The opposite is the case for the \citet{Stevans2018} results, where they find their DPL+DPL fits provide a higher total emissivity. The red and black lines linking our data points and those of \citet{Stevans2018} show the range of emissivities between those found by the Sch+DPL and DPL+DPL parameterisations. The data points for the various observational studies are extracted from Figure 1 of \citet{Madau2015}, who analyses each set of observations in a self-consistent way. The data used originates from \citet{Bongiorno2007,Schulze2009,Gilkman2011,Masters2012,Giallongo2015}. Additional data points from more recent studies have been added from \citet{Giallongo2019}. We also show the results of integrating the galaxy UV LF under the same process but with escape fractions $f_{\rm esc} = 0.1, 1.0$ as the red crosses. We find that the Schechter and DPL functional forms do not cause a significant difference in the estimated emissivity of galaxies.}
    \label{fig:AGNIon}
\end{figure*}

When we examine the luminosity dependence on the evolution of the $k$ parameter, we find there are no significant trends present. This indicates that there is a near uniform rise in the number of AGN between $-26<M_{\rm UV} < -24$ between $4<z<5$. Between $3<z<4$, the growth in number density of AGN slows down significantly. In particular, bright AGN of $M_{\rm UV} =- 26$ are found to be consistent in number density between $z=3$ and $z=4$ while fainter AGN still show growth in their number density. This drives the flattening of the faint-end slope at $z=3$ and shifts of the knee location towards fainter luminosities. Our results are consistent with the evolution proposed in \citet[][see their figure 10]{Niida2020}, who combined a number of studies to explore the luminosity dependent evolution of the AGN LF across a wide range of redshifts \citep[$0.5<z<6$][]{Richards2006,Siana2008,Croom2009,McGreer2013,Akiyama2018,Matsuoka2018c}. Here, they also report that AGN number density rises more rapidly at higher redshifts before their evolution flattens off and eventually turns over, leading to a decrease in the number of UV luminous AGN in the modern Universe. The key finding was that brighter AGN peak in their number density at earlier times than fainter AGN. Those with a luminosity $M_{\rm UV}=-26$ peak around $z=3$ while AGN with $M_{\rm UV}=-24$ peak in number density closer to $z=2$. This is consistent with theoretical predictions of the combined effects of AGN downsizing and dust obscuration \citep[e.g.][]{Fanidakis2012,Hirschmann2014}. 

To explore whether the excess in the number density of bright galaxies at $z\geq7$ \citep[reported in][]{Bowler2014} can be explained by faint AGN, we fit a simple linear model in $(1+z)$ to the value of $k$ and extrapolate to higher redshifts. For this model, we use the $k$ values derived in this study and include those derived for the difference between the $z=5-6$ AGN UV LF in \citet{Niida2020}. At $M_{\rm UV}=-24$, the slope of the evolution in $k$ ($\frac{\delta k}{\delta z}$) is found to be $-0.40\pm0.03$ ($-0.42\pm0.04$) when using the DPL+Sch (DPL+DPL) functional forms. This predicts a value of $k$ between LFs measured at $z=6-7$ to be around $k=-1.3$. Using the $z=6$ UV LF measurements from \citet{Matsuoka2018c}, which have a number density of around $1 \times 10^{-8} {\rm Mpc}^{-3} {\rm mag}^{-1}$,  as a baseline for AGN number densities at $z=6$, we estimate the number density of AGN at $M_{\rm UV}=-24$ will be approximately $5 \times 10^{-10} {\rm Mpc}^{-3} {\rm mag}^{-1}$ based on this trend. This is over two orders of magnitude less than the number density of galaxies observed at $M_{\rm UV}=-23$ in \citet{Bowler2014}. Even with a steep faint-end slope ($\alpha_{\rm AGN} < -2$), the number of AGN at this time is insufficient to explain the excess in number counts reported by \citet{Bowler2014}. Extrapolating this trend to even higher redshift, we would not expect unobscured, Type-I AGN to be significant contaminants of current measurements of the $z\geq7$ galaxy LF. This is in disagreement with recent results from \citet{Leethochawalit2022}, whom claim to observe a substantial number density of sources with $M_{\rm UV} < -22$ at $z=8$. This was attributed to AGN, but the number densities required would be orders of magnitude more than the observed evolution of the AGN UV LF at lower redshifts predict. Our conclusions here are also further supported by the study of \citet{Finkelstein2022}, whose compilation of AGN studies and predictions towards higher redshifts also find that AGN are unable to account for the excess in highly luminous objects found in galaxy LF studies at $z>7$.

\subsection{Ionizing emissivity of AGN in the epoch immediately post-reionization}

A key debate regarding the reionization epoch is the contribution of ionizing photons from AGN compared to the fainter, but more numerous, star-forming galaxies. Evidence has been posed both for  \citep[e.g.][]{Giallongo2015,Madau2015,Yoshiura2017,Bosch2018,Giallongo2019,Dayal2020,Torres2020} and against \citep[e.g.][]{Stark2016,Qin2017,Aloisio2017,Hassan2018,Mitra2018M,Parsa2018,Zeltyn2022} scenarios where AGN provide significant contributions to the reionizing process in the early Universe. Here, we calculate the emissivity of UV photons at 1500\AA\, for the two populations by integrating the fits to our luminosity functions between the luminosity range of $-30<M_{\rm UV}<-15$ [$\rho_{\rm 1500} = \int_{-30}^{-15}\Phi (M) L(M) dM$]. Comparing the result from integrating the galaxy and AGN UV LFs separately, we find that galaxies generate a factor of 40 (100) times more UV (1500\AA) photons than AGN at $z=5$ when using the Schechter (DPL) functional form for the galaxy population. Similarly, galaxies generate 33 (60) times more photons at $z=4$. This shows that the steeper faint-end slopes of the AGN LF found when using the Schechter functional form for the galaxy population results in more significant AGN contributions to the total amount of ultraviolet light at this time. 

The above calculation assumes that AGN and LBGs have the same escape fraction ($f_{esc}$) of ultraviolet photons into the inter-galactic medium, which may not be the case. Some studies have assumed escape fractions of unity for objects classed as AGN \citep{Giallongo2015,Stevans2018} and this is supported by observations that show generally large escape fractions for these sources \citep[$>50$ per cent][]{Cristiani2016,Grazian2018,Romano2019}. However, it is unclear if this trend holds towards fainter luminosities. For galaxies, there are a variety of results that show the escape fraction is dependent on redshift, stellar mass, halo mass and more, with observations measuring escape fractions from the sub-percent level to over 30 per cent \citep[e.g.][]{Cooke2014,Grazian2017,Steidel2018,Vanzella2019,Fletcher2019,Bian2020,Izotov2021,Begley2022}.

To explore the AGN budget of ionizing photons further, we conduct a simple conversion from $\rho_{\rm 1500}$ to $\rho_{\rm 912}$, the emissivity at the Lyman limit, for the AGN population following the same procedure as used in \citet{Stevans2018}. Here, we assume that the spectral slope follows a power law of $\alpha_\nu = -1.41$ between 1000-1500\AA\, \citep{Shull2012} and $\alpha_\nu = -0.83$ for 912-1000\AA\, \citep{Stevans2014}. We note here that \citet{Stevans2018} explores the impact of these SED slope assumptions and find it has an impact of $\sim10$ per cent on the final $\rho_{\rm 912}$ value if different assumptions, based on other observational studies, are used. For closer comparison to other studies that perform this calculation \citep[][]{Madau2015,Stevans2018}, we integrate the UV LF down to $M_{\rm UV} < -18$ when calculating $\rho_{912}$ for AGN. For a simple comparison, we conduct this calculation for galaxies too, assuming an ultraviolet slope of $\alpha_\nu = -1.7$ \citep{Bouwens2014,Wilkins2016} and two values for the escape fraction ($f_{esc}=0.1, 1.0$). The results are presented in Fig. \ref{fig:AGNIon}.

This analysis is once again an optimistic calculation for the contribution of AGN to the number of Hydrogen ionizing photons reaching the neutral IGM due to our assumption of a $f_{esc}=1$. Even with this optimism, we find that the emissivity is below that measured by \citet{Giallongo2015} and subsequently used in the modelling conducted by \citet{Madau2015}. Our estimate is found to be broadly consistent with the results from other studies covering this time period \citep{Gilkman2011,Masters2012,Stevans2018,Giallongo2019} and is greater than a factor of 10 below the emissivity of galaxies when measured under the same assumptions. The use of a Schechter function to model the galaxy population leads to greater numbers of faint-end AGN which results in a larger total emissivity (our downwards pointing triangles) than when using the DPL functional form (our upwards pointing triangles). This is at its most pronounced at $z=3$ and $z=5$, while at $z=4$ the results are comparable due to the consistent fits to the AGN LF that we measure. Within uncertainties, it is still plausible that AGN alone could sustain Hydrogen reionization at $z<5$. Their total contributions to the later stages of the reionization process should not be entirely discounted until more is understood about the number density of very faint AGN ($M_{\rm UV} > -24$) and their leakage of Lyman Continuum photons. With the simple assumptions of $f_{esc}=1.0$ for AGN and $f_{esc}=0.1$ for star-forming galaxies, it is likely that a combination of galaxy and AGN contributions can maintain reionization at lower redshifts. At $z>5$, the extrapolated rapid evolution of the AGN LF observed in this study and others \citep{Matsuoka2018c,Niida2020} will generate expected emissivities that are lower than at $z=5$, indicating that AGN contributions deeper within the reionization epoch at $z>7$ will be relatively small compared to galaxies.

\subsection{The transition between AGN and star-forming dominated UV emission}

\begin{figure}
    \centering
    \includegraphics[width=\columnwidth]{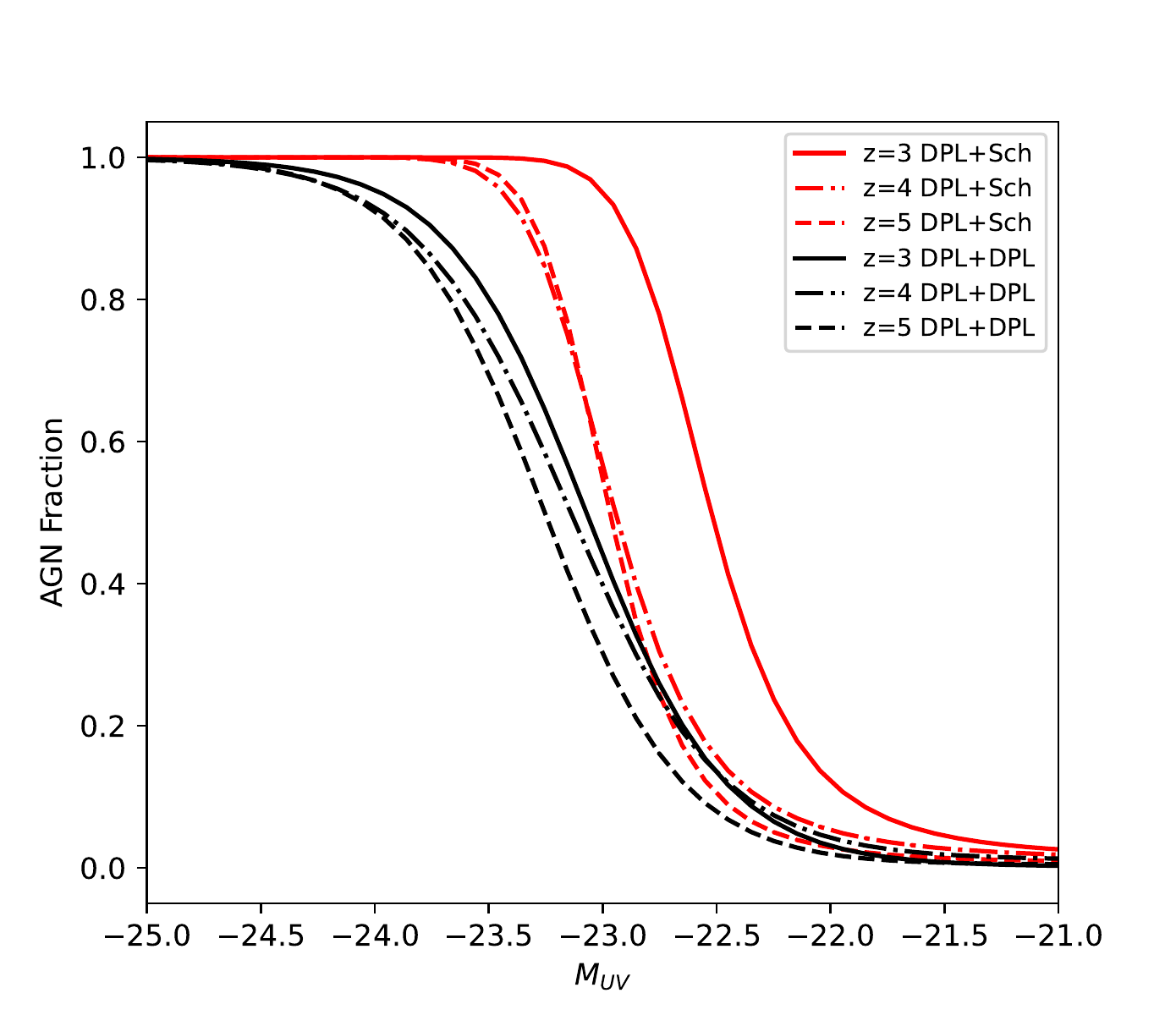}
    \caption{The AGN fraction as inferred by the ratio of the contributions of the double power law (DPL) fit to the AGN population and the combined UV LF of both galaxies and AGN as measured in this study. The red lines are for when a Schechter parameterisation is used to describe the galaxy population and the black lines are for when a DPL parameterisation is used.}
    \label{fig:AGNFrac}
\end{figure}

With a parameterised ultraviolet luminosity function measured for both the AGN population and star-forming galaxies, the question of which population dominates number counts at different absolute luminosities can be explored. In Fig. \ref{fig:AGNFrac}, we show the results of using our measured luminosity functions to calculate the luminosity-dependent AGN fraction in our three redshift bins. Here, we find that the DPL parameterisation for the galaxy UV LF predicts a higher number of ultra-luminous, star-forming galaxies ($M_{\rm UV}<-23$) compared to the Schechter parameterisation. This leads to a more gradual shift between AGN and galaxy-dominated ultraviolet emission (spanning two and a half magnitudes), as opposed to the sharp transition observed when the Schechter parameterisation is used (around one and a half magnitudes).

The use of the DPL functional form for galaxies is found to predict a small evolution in the AGN fraction with redshift. Here, AGN dominance extends towards lower luminosities at lower redshifts. However, current statistical and systematic errors on the faint end of the AGN LF mean this result is presently insignificant. The Schechter functional form, however, exhibits a strong shift between $z\sim4$ to $z\sim3$, brought on by the combination of the rise in faint AGN number counts and the significant fall in the $M^*$ value for the star-forming galaxies across this time period. The cause of this evolution can be attributed to the onset of widespread quenching in massive, luminous galaxies at $z<4$ resulting in the measured turnover in the star-forming main sequence \citep{Muzzin2013,Tomczak2016,Davidzon2017,McLeod2020}.

To answer the question regarding the true number densities of ultra-luminous, star-forming galaxies ($M_{\rm UV}<-22$), the differences in the predicted AGN fractions between our models can be exploited. The region of parameter space where the predictions of the AGN fraction between the DPL or Schechter function are found to differ the most is between $-24<M_{\rm UV}<-23$. Consequently, a highly complete spectroscopic survey targeting a population of sources in this luminosity regime has the potential to shed light on which parameterisation of the luminosity function best describes the intrinsic population. The studies by \citet{Ono2017}, \citet{Boutsia2018} and \citet{Bowler2021} explored this possibility using both new and legacy spectroscopic datasets; however, such datasets present sample sizes that are too small and lacking in completeness for any significant results to presently be determined. In order to successfully make the distinction between the two parameterisations, a targeted spectroscopic survey, perhaps combined with more detailed SED fitting to distinguish the AGN and star-formation components \citep[e.g.][at lower redshifts]{Thorne2021}, is likely required.

\subsection{Evolution of the galaxy UV LF}\label{sec:lfevo}

In contrast to the AGN LF, the evolution in the LF of star-forming galaxies is much more mild across the redshift range of $z=3$ to $z=5$. Compared to the studies undertaken with \emph{Hubble} surveys and the COSMOS field \citep[e.g.][]{Finkelstein2015,Parsa2016,Bouwens2021}, the volumes probed in this study are significantly larger. This enables us to measure the bright end of the galaxy UV LF to much greater precision, placing tight constraints on the bright-end slope and the value of the characteristic luminosity $M^*$. The key finding of this study is that the number density of luminous galaxies $M_{\rm UV}<-21$ increases from $z=5$ to $z=4$, but then decreases between $z\sim4$ to $z\sim3$. This is subsequently observed in the fit parameter $M^*$ in both Schechter and DPL functional forms, which increases at higher redshifts and then decreases at $z\sim3$. 

Evolution in the UV LF of galaxies, and subsequently the parameters used to describe it, is dependent on a number of physical factors that govern galaxy evolution. Star formation rate, quenching mechanisms and dust obscuration all affect its shape and all potentially have dependence on time. The excess of galaxies at $M_{\rm UV}<-22$ present in the more favoured DPL functional fits could challenge our understanding of these processes. Before these can be considered however, the effects of gravitational lensing may need to be taken into account. The studies of \citet{Ono2017}, \citet{Bowler2015,Bowler2020} and \citet{Harikane2021} note that the excess of ultra-luminous galaxies is more than can be attributed to lensing effects from foreground galaxies. To investigate this further, we assume that our measured UV LF is described by a Schechter function that is convolved with a simple lensing model. We conduct this work by following the lensing prescription developed across the studies \citet{Wyithe2011,Barone2015,Ono2017} and \citet{Harikane2021}.

We recalculate the $\chi^2$ using our observed UV LF over the magnitude range $-23<M_{\rm UV}<-20$ using this new lensed Schechter function and compare the results to the initial Schechter and DPL fits to the galaxy population. We find that the predicted number of luminous galaxies with $M_{\rm UV} < -22.5$ increases by only a marginal amount between the original Schechter model and the lensed Schechter model. This leads to small changes in the quality of the fit of $\delta\chi^2\sim1$. This is in agreement with the results of \citet{Ono2017}, who find that strong lensing is most impactful at redshifts higher than probed in this study. This is because the lensing optical depth increases with redshift due to the increasing probability of line-of-sight alignment with foreground, lower-redshift galaxies \citep{Takahasi2011,Barone2015}.

\begin{figure}
    \centering
    \includegraphics[width=\columnwidth]{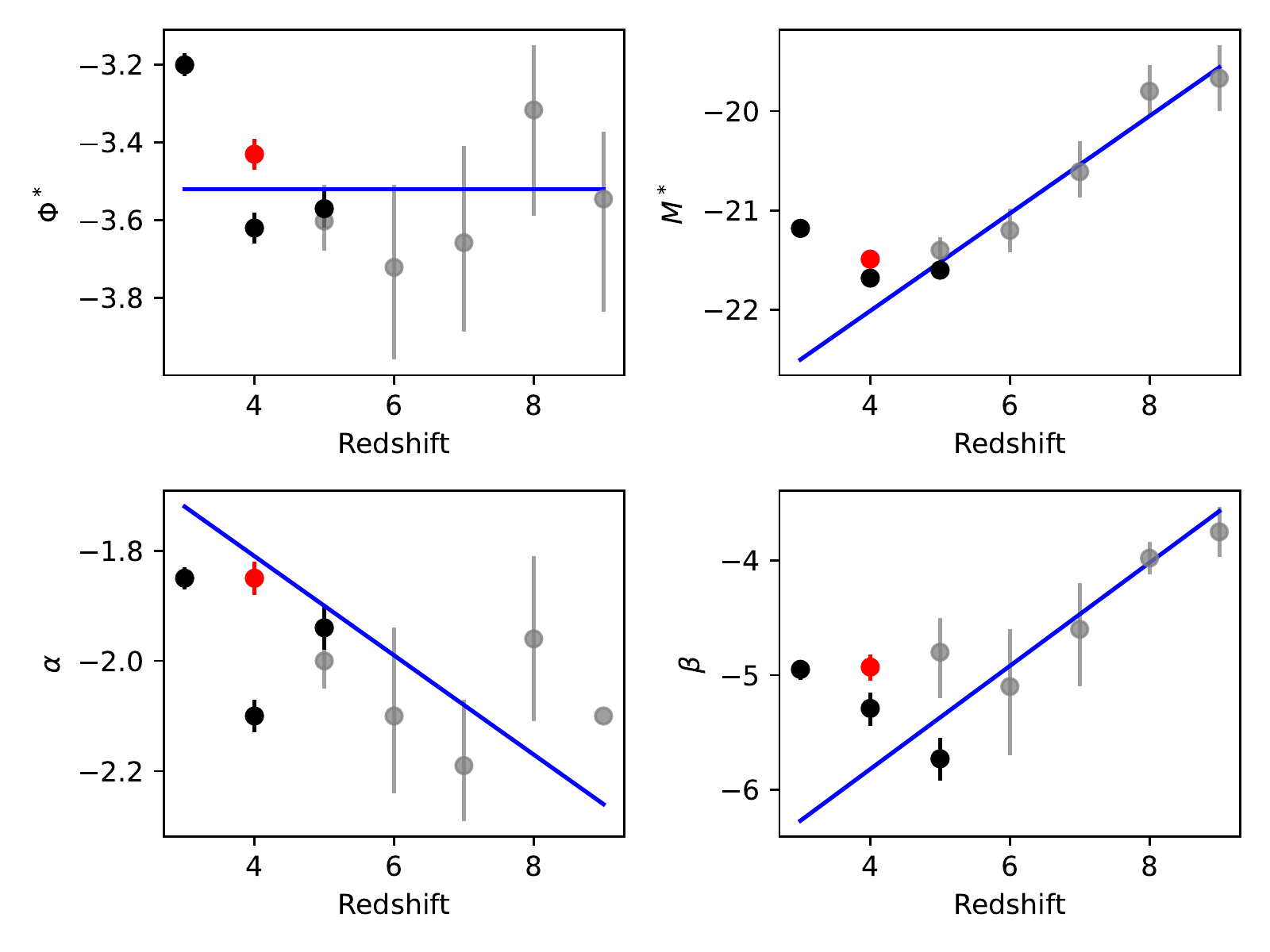}
    \caption{The time evolution of the DPL parameters used to describe the galaxy UV LF. In black we show the results of this study, in lighter gray we show the results from \citet{Bowler2020} and the blue line shows the linear fit to $z=5-8$ conducted in that study. Our $z=5$ UV LF agrees well with the predicted evolution, while at $z=3$ it is clear that some diversion in the behaviour of the LF takes place. The red data points indicates the $z=4$ fit conducted in Appendix \ref{sec:AppB} with the \citet{Finkelstein2015} faint-end data points.}
    \label{fig:BowlerEvo}
\end{figure}

As lensing is unable to explain the presence of these ultra-luminous galaxies, their existence can indicate a lack of quenching and/or dust obscuration at this time. The evolution of the bright-end slope $\beta$ can be used to gain further insight into these processes. The evolution found by \citet{Bowler2020} indicates that $\beta$ steepens with time, keeping the number density of ultra-luminous galaxies at $M_{\rm UV}\sim-23$ nearly constant. They postulate that this gradual steepening of the bright-end slope can be attributed to increased dust content in galaxies towards lower redshifts. As is shown in Figure \ref{fig:BowlerEvo}, our observations agree well with this evolutionary model until $z\simeq3$, where we observe the values of $M^*$ and the bright-end slope ($\beta$) experience a turn over and reversal in their evolution between $3<z<4$. This coincides with the onset of significant fractions of passivity in massive galaxies (stellar mass $\log(M_*/M_\odot) > 10.5$) as time advances from $z\sim4.5$ to $z\sim2.75$, leading to a turnover in the star-forming main sequence at the high-mass end \citep{Muzzin2013,Tomczak2016,Davidzon2017,McLeod2020}. Consequently, the most massive and luminous galaxies see a larger reduction in star formation rates which will ultimately reduce their ultraviolet luminosities. Hence, the impact of dust attenuation may be an important driver in the evolution of the bright end of the UV LF at high redshifts ($z>4$), but quenching of star formation will become increasingly important towards lower redshifts ($z<4$).

\section{Conclusions}\label{sec:conclusions}

In this work, we have presented new measurements of the ultraviolet luminosity function between $2.75<z<5.2$. The sample was selected using a highly complete and robust spectral energy distribution fitting procedure that was applied to $\sim10$\ds of deep optical and near-infrared photometry in the COSMOS, XMM-LSS \& E-CDFS extragalactic fields. These observations were combined with other studies that complement the dynamical range of luminosity and survey volume probed by our study, enabling the ultraviolet luminosity function to be constrained from $-28.5<M_{\rm UV}<-16$ and covering 8 orders of magnitude in number density. The key findings of this study can be summarised as follows:

\begin{enumerate}
    \item The choice of parameterisation used to describe the galaxy population (a DPL or Schechter function) remains difficult to discern by purely photometric means. The $z\sim3$ and $z\sim5$ UV LFs shows a strong preference for a DPL functional form. However, the results from $z\sim4$ show the qualities of fits remain too close to favour one over the other. A simple lensing model is found to be unable to significantly improve the quality of the Schechter-based fits. 
    \item The number density of ultra-luminous galaxies at $z\sim3$ is lower than that of $z\sim4$ and fit values of $M^*$ decrease dramatically between $3<z<4$. This result supports studies that find an earlier turnover in the value of $M^*$ \citep{Weisz2014,Parsa2016}. Comparisons to simple evolutionary models from \citet{Parsa2016,Bowler2020} and \citet{Bouwens2021} indicate that $z\sim4$ is a key time period where the evolution of the UV LF experiences a reversal in the growth in the number density of the most UV-luminous galaxies. This is likely due to more widespread quenching in the population of massive galaxies between $2.75<z<4.5$ \citep{Muzzin2013,Tomczak2016,Davidzon2017,McLeod2020}, indicating that dust attenuation may drive the evolution of the bright end of the galaxy UV LF at high redshifts \citep[$z>4$,][]{Bowler2020} while the quenching of star formation becomes more impactful at lower redshfits ($z<4$).
    \item Utilising the differing predictions in the AGN fraction with absolute UV luminosity is required to solve the above issue in discerning which of the Schechter or DPL parameterisations better describes the galaxy population. We find that a spectroscopic campaign targeting objects with an absolute luminosity of $-24<M_{\rm UV}<-23.25$, expanding upon work started by \citet{Boutsia2018,Ono2017} and \citet{Bowler2021}, will provide a definitive answer to the number densities of UV-faint AGN and UV-luminous star-forming galaxies at this time.
    \item Examining the time evolution of the AGN fraction versus $M_{\rm UV}$, we find that using the DPL parameterisation to model the galaxy population produces a non-evolving AGN fraction, with AGN dominating number counts at luminosities brighter than $M_{\rm UV}<-23.1$. On the other hand, using the Schechter parameterisation to model the galaxies provides a much sharper evolution, with AGN domination extending as faint as $M_{\rm UV} = -22.6$ at $z=3$.
    \item The number density of UV-faint AGN is found to evolve much more rapidly at earlier times and grows by around two orders of magnitude between $3<z<6$. This indicates a rapid onset of AGN activity as the Universe enters its current, reionized state. Current data quality is presently unable to discern significant luminosity dependence in the rise in number counts of UV-faint AGN between $z=5$ and $z=3$, though modelling of the wider LF shows the faint-end slope of the AGN LF flattens with time. Extending our measured evolution of the AGN LF to $z\sim7$, we find the predicted number densities of UV-faint AGN to be too few to account for the excess in galaxies identified in \citet{Bowler2014,Bowler2020}.
    \item Following an optimistic calculation for the AGN contributions towards reionizing photons, we obtain values for $\rho_{\rm 912}$ that are lower than found by \citet{Giallongo2015} and the model derived in \citet{Madau2015}, indicating that AGN contributions are not high enough to be a dominant contributor towards maintaining Hydrogen reionization. Although, with our measured emissivities, we find that AGN contributions are high enough to not be discounted entirely. This is especially the case if galaxies are better described with a Schechter function as opposed to a DPL functional form due to the higher numbers of AGN that are required to match the observed number density of sources at $M_{\rm UV}\sim-23$. The rapid evolution of the AGN LF makes it increasingly unlikely that AGN contribute significantly to the initial reionization of Hydrogen at $z>7$.
\end{enumerate}

Up until now, studies that probe the very faint end of the UV LF ($M_{\rm UV}>-20$) at high redshifts have primarily been focused on using the \emph{Hubble} Space Telescope, which as a relatively small field of view and near-infrared coverage limited to 1.8 microns.  This has restricted the wavelength ranges available to apply more detailed SED modelling and redshift estimation as used in this work. With the increase in depths, wavelength coverage and volume probed by upcoming facilities such as the \emph{James Webb Space Telescope} (JWST) and the upcoming \emph{Euclid} Space Telescope, the application of SED fitting techniques will be able to be extended to even higher redshifts and larger dynamical ranges of intrinsic luminosity. The resulting improvement in both completeness and contamination rates that photometric redshifts/SED modelling provide over colour-colour selection \citep[e.g.][]{vdb2010,Oesch2013} will soon enable for greater constraints to be placed on the total UV LF across most of the observable Universe. 

\section{Data Availability}
All imaging data was obtained from original sources in the public domain. Associated references to each survey utilised are provided within Section \ref{sec:data} and information on how to obtain the data are contained therein. Catalogues containing photometry and redshifts have been published through \citet{Hatfield2022}. The specific sample used in the measurement of the UV LF in this study is provided as online material to accompany this work.

\section*{Acknowledgements}

The authors would like to pass on our thanks to J.Patterson and the University of Oxford's IT team at the Physics Department for their continued efforts. This research  made  use  of Astropy,  a  community-developed core  Python  package  for  Astronomy  (Astropy  Collaboration,  2013).

NA acknowledges funding from the Science and Technology Facilities Council (STFC) Grant Code ST/R505006/1 and support from the European Research Council (ERC) Advanced Investigator Grant
EPOCHS (788113). RB acknowledges support from an STFC Ernest Rutherford Fellowship [grant number ST/T003596/1]. RGV acknowledges funding from the Science and Technology Facilities Council (STFC) [grant code ST/W507726/1]. MJJ acknowledges support of the STFC consolidated grant [ST/S000488/1] and
[ST/W000903/1] and from a UKRI Frontiers Research Grant [EP/X026639/1]. MJJ also acknowledges support from
the Oxford Hintze Centre for Astrophysical Surveys which is funded
through generous support from the Hintze Family Charitable Foundation. This work was supported by the Glasstone Foundation and the award of the STFC consolidated grant (ST/N000919/1). 

This work is based on data products from observations made with ESO Telescopes at the La Silla Paranal Observatory under ESO programme ID 179.A-2005 and ID 179.A-2006 and on data products produced by CALET and the Cambridge Astronomy Survey Unit on behalf of the UltraVISTA and VIDEO consortia.

Based on observations obtained with MegaPrime/MegaCam, a joint project of CFHT and CEA/IRFU, at the Canada-France-Hawaii Telescope (CFHT) which is operated by the National Research Council (NRC) of Canada, the Institut National des Science de l'Univers of the Centre National de la Recherche Scientifique (CNRS) of France, and the University of Hawaii. This work is based in part on data products produced at Terapix available at the Canadian Astronomy Data Centre as part of the Canada-France-Hawaii Telescope Legacy Survey, a collaborative project of NRC and CNRS.

The Hyper Suprime-Cam (HSC) collaboration includes the astronomical communities of Japan and Taiwan, and Princeton University. The HSC instrumentation and software were developed by the National Astronomical Observatory of Japan (NAOJ), the Kavli Institute for the Physics and Mathematics of the Universe (Kavli IPMU), the University of Tokyo, the High Energy Accelerator Research Organization (KEK), the Academia Sinica Institute for Astronomy and Astrophysics in Taiwan (ASIAA), and Princeton University. Funding was contributed by the FIRST program from Japanese Cabinet Office, the Ministry of Education, Culture, Sports, Science and Technology (MEXT), the Japan Society for the Promotion of Science (JSPS), Japan Science and Technology Agency (JST), the Toray Science Foundation, NAOJ, Kavli IPMU, KEK, ASIAA, and Princeton University.

This paper is based, in part, on data collected at the Subaru Telescope and retrieved from the HSC data archive system, which is operated by Subaru Telescope and Astronomy Data Center at National Astronomical Observatory of Japan. Data analysis was in part carried out with the cooperation of Center for Computational Astrophysics, National Astronomical Observatory of Japan.




\bibliographystyle{mnras}
\bibliography{mnras_template} 


\appendix
\section{Tabular data set}
\begin{table}
\caption{The measured rest-frame UV LF and its error margin at $2.75<z<3.5$, $3.5<z<4.5$ and $4.5<z<5.2$. Column 1 shows the absolute UV magnitude at 1500\AA. Column 2 shows the number density of objects and column 3 shows the errors in the number density which are calculated with equation 2 and summed in quadrature with the derived cosmic variance from Section\ref{sec:CV}.}
\centering
\begin{tabular}{lll}
\hline
$M_{\rm UV}$     & $\Phi (10^{-4})$                        & $\delta\Phi (10^{-4})$                  \\
$[\textrm{mag}]$ & $[\textrm{mag}^{-1} \textrm{Mpc}^{-3}]$ & $[\textrm{mag}^{-1} \textrm{Mpc}^{-3}]$ \\ \hline
$z=3.1$          &                                         &                                         \\
-26.375 & 0.00017 & 0.00015 \\
-25.625 & 0.00143 & 0.00050 \\
-24.875 & 0.00278 & 0.00064 \\
-24.125 & 0.00528 & 0.00091 \\
-23.375 & 0.00802 & 0.00111 \\
-22.875 & 0.01974 & 0.00319 \\
-22.625 & 0.03701 & 0.00427 \\
-22.375 & 0.08352 & 0.00668 \\
-22.125 & 0.19731 & 0.01109 \\
-21.875 & 0.47466 & 0.02032 \\
-21.625 & 0.99727 & 0.03651 \\
-21.375 & 1.99722 & 0.06680 \\
-21.125 & 3.46469 & 0.11091 \\
-20.875 & 5.58575 & 0.17453 \\
-20.625 & 8.16910 & 0.25193 \\
-20.375 & 11.3134 & 0.3461 \\
-20.125 & 13.8824 & 0.4300 \\
-19.875 & 17.9392 & 0.6388 \\
-19.625 & 20.4542 & 0.8361 \\
-19.375 & 23.4597 & 0.9553 \\ \hline
$z=4.0$          &                                         &                                         \\
-26.375 & 0.00014 & 0.00013 \\
-25.625 & 0.00037 & 0.00025 \\
-24.875 & 0.00112 & 0.00040 \\
-24.125 & 0.00199 & 0.00050 \\
-23.500 & 0.00563 & 0.00113 \\
-23.125 & 0.01035 & 0.00216 \\
-22.875 & 0.03118 & 0.00388 \\
-22.625 & 0.06534 & 0.00571 \\
-22.375 & 0.16073 & 0.00992 \\
-22.125 & 0.33717 & 0.01667 \\
-21.875 & 0.80920 & 0.03369 \\
-21.625 & 1.49534 & 0.05788 \\
-21.375 & 2.10349 & 0.09054 \\
-21.125 & 3.23644 & 0.13577 \\
-20.875 & 5.09635 & 0.23695 \\
-20.625 & 7.25133 & 0.33381 \\
-20.375 & 9.97100 & 0.45609 \\
-20.125 & 11.8694 & 0.5420 \\ \hline
$z=4.8$          &                                         &                                         \\
-25.000 & 0.00016 & 0.00015 \\
-24.000 & 0.00161 & 0.00048 \\
-23.250 & 0.00459 & 0.00114 \\
-22.875 & 0.00946 & 0.00228 \\
-22.625 & 0.03599 & 0.00470 \\
-22.375 & 0.09267 & 0.00837 \\
-22.125 & 0.24209 & 0.02030 \\
-21.875 & 0.60223 & 0.04240 \\
-21.625 & 1.19789 & 0.07810 \\
-21.375 & 2.28618 & 0.15824 \\
-21.125 & 3.39954 & 0.32140 \\
-20.875 & 5.30760 & 0.49300 \\
-20.625 & 7.02957 & 0.64748 \\\hline
\end{tabular} \label{Tab:Points}
\end{table}

Presented in Table~\ref{Tab:Points} are the binned rest-frame UV LF data points for each of the three redshift bins explored in this study $2.75 < z < 3.5$, $3.5 < z < 4.5$ and $4.5 < z < 5.2$.


\section{Alternate $z=4$ UV LF fit}\label{sec:AppB}

\begin{table*}
\centering
\caption{The results of the MCMC fitting applied to the total UV LF at $z=4$. The fit uses data points from \citet{Finkelstein2015} fainter than probed by our measurement. It also includes those from \citet{Akiyama2018} to provide better sampling of the AGN UV LF.
The first column lists the fitting parameterisation used. Columns 2--5 are the best fit DPL parameters for the AGN LF. Columns 6--9 show the best fit LBG parameters, either for a Schechter function or DPL.
The final columns show the $\chi^2$ and reduced $\chi^2$ of the fit.}\label{tab:Result4}
\begin{tabular}{l|llllllll|ll}
\hline
Function  & $\textrm{log}_{10}(\Phi_{\text{AGN}})$ & $M^*_{\text{AGN}}$ & $\alpha_{\text{AGN}}$ & $\beta_{\text{AGN}}$ & $\textrm{log}_{10}(\Phi)$ & $M^*$ & $\alpha$ & $\beta$ & $\chi^2$ & $\chi_{\rm red}^2$ \\
  & $\textrm{mag}^{-1} \textrm{Mpc}^{-3}$ & mag &  &  & $\textrm{mag}^{-1} \textrm{Mpc}^{-3}$ & mag &  &  &  &  \\\hline
$z=4.0$        &                     &                           &       &          &         &          & &  & &                     \\
DPL+Sch               & $-7.96^{+0.13}_{-0.10}$   & $-27.41^{+0.16}_{-0.12}$ & $-2.14^{+0.05}_{-0.04}$ &  $-4.83^{+0.45}_{-0.55}$                     &  $-2.87^{+0.02}_{-0.03}$ & $-20.92^{+0.03}_{-0.03}$   & $-1.49^{+0.03}_{-0.03}$      &  ---        &  78.11        &  2.17                  \\
DPL+DPL               & $-7.67^{+0.23}_{-0.18}$   & $-27.05^{+0.39}_{-0.24}$ & $-1.95^{+0.13}_{-0.09}$ &  $-4.13^{+0.42}_{-0.49}$                     &  $-3.43^{+0.04}_{-0.04}$ & $-21.49^{+0.05}_{-0.05}$   & $-1.85^{+0.03}_{-0.03}$      &  $-4.93^{+0.11}_{-0.12}$         &  111.05       &  3.17               \\ \hline

\end{tabular}
\end{table*}

Our measured UV LF at $z=4$ is found to have the greatest discontinuity with the \citet{Bouwens2021} data set at the very faint end of the galaxy UV LF. In addition, it is in this redshift bin where the \citet{Bouwens2021} observations differs the most from other studies that have probed the faint end between $3<z<5$ \citep[e.g.][]{Finkelstein2015,Parsa2016}. To assess the potential impact of this on our conclusions regarding the DPL-based evolutionary model discussed in Section \ref{sec:lfevo}, we repeat the fitting procedure with the use of the \citet{Finkelstein2015} data points instead of \citet{Bouwens2021}. We present the results of this in Table \ref{tab:Result4} and highlight it as the red data point in Figure \ref{fig:BowlerEvo}. High $\chi^2$ values remain around the region where we transition from using our observations to that of \citet{Finkelstein2015}, leading to no improvement in the final $\chi^2_{\rm red}$. A possible solution would be to consistently analyse both ground-based and space-based data with the same selection procedures and methodologies. 

We observe that the change in the best-fit parameters is focused on the faint end slope ($\alpha$) of the galaxy UV LF. We find that the LF normalisation $\Phi^*$ increases by around $0.1-0.2$ dex and the characteristic magnitude gets fainter by $0.19$ magnitudes compared to when \citet{Bouwens2021} data is used for both the Schechter and DPL parameterisations. In both cases, these shifts are driven by degeneracies of these parameters with the faint-end slope, which shifts by $\delta\alpha>0.25$. This result also provides a more smoothly evolving DPL faint-end slope compared to when \citet{Bouwens2021} data is used at $z=4$. Otherwise, the results still agree with our observations that the evolution in the parameters used to describe the UV LF experience a turn over around $3<z<4$, indicating a shift in the physical processes governing the evolution of the UV LF, such as the onset of wide scale galaxy quenching.




\bsp	
\label{lastpage}
\end{document}